\documentclass[lettersize,journal]{IEEEtran}

\usepackage{amsmath,amsfonts}
\usepackage{array}
\usepackage[caption=false,font=normalsize,labelfont=sf,textfont=sf]{subfig}

\usepackage{listings}
\usepackage[dvipsnames]{xcolor}

\usepackage{textcomp}
\usepackage{stfloats}
\usepackage{url}
\usepackage{verbatim}
\usepackage{graphicx}
\usepackage{cite}
\usepackage{amssymb}

\usepackage{color}
\usepackage{xcolor}
\usepackage{makeidx}
\usepackage{pifont}
\usepackage{graphicx,floatrow}
\usepackage{stfloats}
\usepackage{epsfig}
\usepackage{epstopdf}
\usepackage{multirow}
\usepackage{multicol}
\usepackage{array}
\usepackage{enumerate}
\usepackage{paralist}
\usepackage{CJK}
\usepackage{url}
\usepackage{float}
\graphicspath{{figs/}}
\usepackage{bm}
\usepackage{makecell}
\usepackage{cite}

\makeatletter  
\newif\if@restonecol  
\makeatother

\usepackage[linesnumbered,ruled,vlined]{algorithm2e}%[ruled,vlined]{
\usepackage{algpseudocode}  
\usepackage{amsmath}  
\graphicspath{{figs/}}

\newcommand{\bi}{\begin{itemize}}
\newcommand{\ei}{\end{itemize}}
\newcommand{\be}{\begin{enumerate}}
\newcommand{\ee}{\end{enumerate}}

\newcommand{\eg}{{e.g., }}

\newcommand{\ynote}[1]{}

\newcommand{\oldstuff}[1]{}
\newcommand{\info}[1]{}
\newcommand{\old}[1]{}
\newcommand{\optional}[1]{}
\newcommand{\consider}[1]{}
\newcommand{\moved}[1]{}
\newcommand{\comments}[1]{}
\newcommand{\temp}[1]{}

\newcommand{\revision}[1] {\textcolor{black}{#1}}
{\begin{list}{$\bullet$}{\setlength{\leftmargin}{1.5ex}%  originally 1 ex
\setlength{\itemindent}{.5ex}}}% originally .5
{\end{list}}

\DeclareUnicodeCharacter{2061}{}

\ifCLASSINFOpdf
\else
\fi
\begin{document}

\title{CoAvoid: Secure, Privacy-Preserved Tracing of Contacts for Infectious Diseases}

% \author{Teng Li\orcidlink{0000-0001-5147-8336}, Siwei Yin\orcidlink{0000-0002-1679-2706}, Runze Yu\orcidlink{0000-0003-1767-571X}, Yebo Feng\orcidlink{0000-0002-7235-2377}, Lei Jiao\orcidlink{0000-0002-3964-3172}, Yulong Shen, and Jianfeng Ma 

\author{Teng Li, Siwei Yin, Runze Yu, Yebo Feng, Lei Jiao, Yulong Shen, and Jianfeng Ma

\thanks{Teng Li, Siwei Yin, Runze Yu, and Jianfeng Ma are with the School of Cyber Engineering, Xidian University, Shaanxi, China. Email: litengxidian@gmail.com, 21151213627@stu.xidian.edu.cn, mercy2green@gmail.com, and jfma@mail.xidian.edu.cn.

Yebo Feng and Lei Jiao are with the Computer and Information Science Department, University of Oregon, USA. Email: \{yebof, jiao\}@cs.uoregon.edu.

Yulong Shen is with the School of Computer Science, Xidian University, Shaanxi, China. Email: ylshen@mail.xidian.edu.cn.
}%
% \thanks{Manuscript received Jan 15, 2022; revised May 2, 2022.}
}

% \markboth{IEEE Journal on selected areas in communications,~Vol.~X, No.~X, XXXX~2022}%
% {Li \MakeLowercase{\textit{et al.}}: CoAvoid: Secure, Privacy-Preserved Tracing of Contacts for Infectious Diseases}

\maketitle

\begin{abstract}

To fight against infectious diseases (\eg SARS, COVID-19, Ebola, etc.), government agencies, technology companies and health institutes have launched various contact tracing approaches to identify and notify the people exposed to infection sources.
However, existing tracing approaches can lead to severe privacy and security concerns, thereby preventing their secure and widespread use among communities. To tackle these problems, this paper proposes \texttt{CoAvoid}, \revision{an edge-based}, privacy-preserved contact tracing system that features good dependability and usability. \texttt{CoAvoid} leverages the Google/Apple Exposure Notification (GAEN) API to achieve decent device compatibility and operating efficiency.
It utilizes Bluetooth Low Energy (BLE) to detect close contact with other people and leverages GPS with fine-grained matching algorithms to verify user information.
In addition, to enhance privacy protection, \texttt{CoAvoid} applies fuzzification and obfuscation measures to shelter sensitive data, making both servers and users agnostic to information of both low and high-risk populations. The evaluation demonstrates good efficacy and security of CoAvoid.
\revision{Compared with four state-of-the-art contact tracing applications, \texttt{CoAvoid} can reduce the size of upload data by at least 90\% and reduce the verification time by 92\%. More importantly, \texttt{CoAvoid} can preserve user privacy and resist replay and wormhole attacks in all analysis scenarios. 
}
\end{abstract}


\section{Introduction}
\IEEEPARstart{C}{ontact} tracing is an effective approach to curb epidemics, letting people know that they may have been exposed to a certain infectious disease (\eg SARS, COVID-19, Ebola, etc.). It can remind high-risk groups to take immediate medical or quarantine measures, thereby interrupting chains of disease transmission~\cite{whocontacttracing}. Typically, a contact tracing application records contact histories between users and leverages information of confirmed patients to determine whether a user is at risk of infection.

Ever since the establishment of the modern public health system, contact tracing has been widely studied. From tracking through questionnaires~\cite{satin1977record} in the 20th century, it has gradually evolved into tracking using digital mobile devices after 2010~\cite{prasad2017spice, firestone2012adding,al2016infection}. From 2020, to reduce COVID-19-associated mortality, various tracing approaches have emerged~\cite{9314850,9519581,9333939}. For example, BlueTrace~\cite{bay2020bluetrace} determines high-risk groups by collecting and analyzing patient information in a central server, which operates effectively, but may raise severe privacy concerns. On the other hand, Epione~\cite{2004.13293} utilizes Private Set Interaction Cardinality to conduct tracing; Whisper~\cite{loiseau2020whisper} utilizes BLE to locally exchange anonymous and temporary identities. These approaches can protect user privacy to a certain extent. However, neither Whisper nor Epione can resist replay attacks~\cite{chowdhury2020covid}.

To standardize and promote developments of contact tracing applications on mobile platforms, in April 2020, the Google/Apple Exposure Notification (GAEN) API, a contact tracing development tool based on Bluetooth Low Energy (BLE), was proposed by joint efforts of Google and Apple~\cite{GoogleAP}. The launch of GAEN has vitally evolved the design of contact tracing applications, as it provides broad hardware and software (IOS and Android) compatibility. Therefore, most state-of-art contact tracing applications are based upon GAEN. Nonetheless, due to inappropriate uses of GAEN API, many GAEN-based applications have two major drawbacks regarding privacy and security of contact tracing, preventing their secure and widespread use among communities~\cite{2004.03544,cryptoeprint:2020:531,9488728}. First of all, such applications tend to expose all the information of confirmed patients to severs and relevant users~\cite{gvili2020security,9144194}, which enables some entities to gather multiple information about patients to infer further their identities, daily routines, or even social relationships~\cite{ji2015secgraph,radaelli2018quantifying}. Besides, as a limitation of Bluetooth, applications that only rely on GAEN can only approximately estimate the distance between users~\cite{cunche2020using,hatke2020using}. This flaw enables attackers to interfere with a user's Bluetooth device to carry out wormhole attacks~\cite{baumgartner2020mind}. Consequently, users can receive an excess of false alarms, putting a severe strain on the public health system and causing social panic.

\IEEEpubidadjcol
As per the aforementioned missing gaps, we seek to
preserve the privacy protection of users while tracing their contacts, which is not only limited to the low-risk population, but also to prevent confirmed patients’ identities and social relationships from disclosing. Additionally, the tracing system should feature comprehensive security in its operations, generating legitimate outputs even in the absence of wormhole and replay attacks. Furthermore, to operate in diverse network and device environments, the proposed approach ought to have high efficiency and board compatibility.

To solve the above issues, we designed \texttt{CoAvoid}, an edge-based contact tracing system that can efficiently track and protect users' security and privacy.
Based upon the GAEN API, \texttt{CoAvoid} can be deployed in both IOS and Android systems equipped with Bluetooth and GPS, and this good compatibility is vital for deployment in a large scale.
To safeguard the tracing system, \texttt{CoAvoid} harnesses several techniques in operations: (1) All the nodes record and verify timestamps along with contact information, thereby eradicating the threat of replay attacks. (2) More than just obtaining relative location via BLE, \texttt{CoAvoid} also collects the user's geographic location via GPS and reliably verifies the user's location with a reliable lightweight algorithm that enables the system to resist wormhole attacks. Furthermore, \texttt{CoAvoid} enhances privacy protection for users from multiple aspects: (1) Every node of \texttt{CoAvoid} system shelters its GPS coordinates through hashing and fuzzification. (2) Confirmed patients filter out uncorrelated contact data and only broadcast essential information to servers. (3) The servers obfuscate information uploaded by confirmed patients before storage and analysis. (4) \texttt{CoAvoid} collects, processes, and stores all the user information anonymously. Due to this nature, \texttt{CoAvoid} is thereby general data protection regulation (GDPR)~\cite{regulation2018general} and California Consumer Privacy Act (CCPA)~\cite{de2018guide} compliant.
Consequently, neither patient's nor ordinary people's identities, daily routines, or social relationships can be leaked or inferred by any entity inside or outside \texttt{CoAvoid} system. Meanwhile, as the data users need to upload and download has been significantly reduced, \texttt{CoAvoid} can achieve a rapid velocity in data transmission.

Compared with existing contact tracing approaches, \texttt{CoAvoid} makes the following contributions:

\begin{itemize}
    \item \revision{While accurately conducting contact tracing, \texttt{CoAvoid} can still protect privacy of both low and high-risk populations in a practical and efficient way. This was almost unreachable through previous approaches, as they face a conflict between data integrity and privacy preserving.}
    \item Benefits from comprehensive security protection designs, \texttt{CoAvoid} can resist replay and wormhole attack in extreme scenarios, enabling the tracing system to properly operate and contact data to be securely transmitted in the absence of malicious users.
    \item By reducing the amount of data required for uploading and analysis, \texttt{CoAvoid} can significantly increase the operating efficiency. With lower bandwidth, storage space, and device performance requirements, \texttt{CoAvoid} can be deployed in regions with different levels of development, promoting widespread use of contact tracing applications.
\end{itemize}

The evaluation results also demonstrate \texttt{CoAvoid}'s enhanced privacy protection in processing users' contact histories, high efficacy in determining high-risk populations, comprehensive security in system operations, and rapid velocity in data transmission. It can reach a 100\% accuracy in contact tracing, with the amount of patient-upload data reduced by 90\% and verification time reduced by 92\%. More importantly, \texttt{CoAvoid} can preserve user privacy, and resist replay and wormhole attacks in all the analysis scenarios.

The rest of this paper is organized as follows. After we
outline related work in Section~\ref{sec:related}, we describe the operation \& threat model of contact tracing in Section~\ref{sec:motivation}.
Then, we elaborate on the system design of the proposed approach in Section~\ref{sec:design}, analyze and evaluate it in Section~\ref{sec:analysis}, and conclude the paper in Section~\ref{sec:concl}.
\section{Related work}
\label{sec:related}

Contact tracing is a vital control tool for infectious diseases and has been developed for decades, with contact tracing available for diseases such as tuberculosis and HIV. In the early stages, when smart devices were not widely available, health care workers collected information through questionnaires to determine people with possible diseases~\cite{satin1977record}. Later, as technology evolved, tracing through portable computing devices has emerged. For example, EbolaTracks~\cite{tracey2015ebolatracks} is a short message service (SMS)-based system for monitoring people who may have been exposed to EVD, including travelers returning from Ebola-affected countries. Outbreaks Near Me~\cite{freifeld2010participatory} work together as a large disease information exchange platform by users uploading disease-related information on their own. However, the information submitted is mixed and data-intensive, and verification of the information is often difficult. It is also difficult to manage the notification of the public while protecting the privacy of those involved in the event. ENACT~\cite{prasad2017enact} detects whether two users are in contact through WiFi signal strength. ENACT dynamically scans the user's surroundings for wireless signals and access points and logs them into the phone. The patient sends this information to the server, alerting other users.

In 2020, the outbreak of COVID-19 led to widespread interest and the rapid development of contact tracing technology. A variety of methods for contact tracing have emerged, including magnetic fields~\cite{bian2020social}, NFC~\cite{10.1007/978-3-030-73216-5_26}, IOT~\cite{9475175}, WiFi~\cite{li2021vcontact,trivedi2021wifitrace}, with Bluetooth Low Energy (BLE)~\cite{9464051} is the most popular method because of privacy and granularity issues. In the initial design of Ad Hoc networks, BLE is widely used for distance measurement and indoor positioning~\cite{10.1145/3143361.3143385,2018BLE,GuoXiansheng2020A}. It has the characteristics of low power consumption, fast connection speed and long distance. By reading the receiver's Received Signal Strength Indicator (RSSI), the distance between the receiving device and the transmitting device can be measured, enabling the convenient contact tracing application to measure whether there is contact between two users. However, range measurement based on RSSI can be affected by many factors such as environment, transmitting power, receiver sensitivity and so on. BTrack~\cite{2018BTrack} introduced a new positioning system, which combined the information of barometers and accelerometers to estimate the user's position. In addition, NOVID ~\cite{loh2020accuracy} uses a combination of Bluetooth and ultrasound and is calculated using sound propagation time. Radio Frequency Identification (RFID) technology can also detect intimacy and social interaction. Suppose an RFID reader is placed in each house room, and everyone carries an RFID tag. The reader can then locate the location by detecting the tag, but this method is not feasible outside and is too expensive. In addition, there are ways to keep track of users. In China, for example, the health code system~\cite{mozur2020coronavirus} is widely used, and it is an epidemic prevention measure based on mobile phones, 3-D face recognition, and population management on multiple occasions.

BlueTrace~\cite{bay2020bluetrace} was one of the first proposed digital contact tracing protocols based on a centralized architecture. This protocol developed the TraceTogether~\cite{stevens2020tracetogether} in Singapore and the Covid-Safe application in Australia. Another protocol called ROBERT was proposed by Inria and Fraunhofer AIESEC~\cite{castelluccia:hal-02611265}. However, contact tracing applications based on a centralized framework collect a large amount of personal information about users. The back-end server can uniquely associate anonymous identifiers with each user and monitor users comprehensively. Therefore, due to privacy and security concerns, all contact tracing applications developed today use a decentralized architecture~\cite{ali2021study,sowmiya2021survey}. Altuwaiyan et al.~\cite{altuwaiyan2018epic} developed an EPIC system based on smartphones, servers, and short-range wireless devices. However, this method collects and even releases personal privacy data directly, which may violate the privacy of individuals.

Since then, contact tracing apps have increasingly focused on protecting users' privacy~\cite{Pinkas2021HashomerP}. Epione~\cite{2004.13293}, for example, utilizes a private set interaction base (PSI-CA) with strong privacy guarantees. When a user has tested positive, the patient's token is encrypted by the server's public key and then sent to Epione's server. Other users use PSI to compare their own token sets with patients' tokens.
Whisper~\cite{loiseau2020whisper} uses BLE to exchange locally generated anonymous and temporary secure identities. However, neither Epione nor Whisper takes into account replay attacks. 
The Delayed Authentication mechanism~\cite{pietrzak2020delayed} may be applied to prevent such attacks after infected people publish temporary identities, but it also brings a new problem on undeniable evidence. Serge Vaudenay~\cite{cryptoeprint:2020:531} has proposed a protocol that uses two-way interaction rather than a broadcast model to mitigate such attacks. The model requires a challenge-and-response protocol between the broadcast device and the receiving device to realize bidirectional communication. The model fundamentally changes the communication architecture of the system, which can lead to various other problems such as more power consumption or more communication problems in complex systems.
Desire~\cite{castelluccia2020desire} is one of the example protocols that follow a hybrid architecture, where the server is responsible for performing the risk analysis and notification process while the client manages the generation of temporary identities. Desire creates and stores a Diffie-Hellman key for each contact between two devices exchange, posesing a high hurdle for resource-constrained devices.

Recently, Google and Apple have collaborated to develop the GAEN API~\cite{GoogleAP}, which provides better device compatibility and is supported by a wide range of device hardware, making this API widely used by recent contact tracing applications, such as SwissCOVID in Switzerland~\cite{SwissCOVID}, Immuni~\cite{immuni} in Italy, and COVIDWISE~\cite{COVIDWISE} in Virginia. However, Baumgärtner et al.~\cite{baumgartner2020mind} pointed out proved experimentally that GAEN design is vulnerable to profiling and possibly de-anonymizing infected persons, and wormhole attacks that principally can generate fake contacts with the potential of significantly affecting the accuracy of the contact tracing system.

    \begin{table}[htp]
    	\renewcommand\arraystretch{1.3}    
    	\tabcolsep=2pt
    	\footnotesize
    	\begin{floatrow}
    		\ttabbox{}{%
    			\begin{tabular}{lcccccc}  
    				\Xhline{1.2pt}		
    				Approaches              &Immuni          &DP-3T         &COVIDWISE   &TraceTogether           &CoAvoid                   \\\Xhline{1.2pt}   		
    				Privacy Protection      &\ding{55}       &\ding{55}     &\ding{55}   &\ding{51}                &\ding{51} \\ 
    				Attack Prevention       &\ding{55}       &\ding{55}     &\ding{55}   &\ding{55}                 &\ding{51}  \\ 
    				Data Upload             &DTK             &DTK           &All         &All                       &RPI   \\ 
    				Verification Rate	    &Slow           &Slow          &Slow         &Medium                     &Fast   \\   %		
    				GAEN-based	            &\ding{51}      &\ding{55}     &\ding{51}    &\ding{55}             &\ding{51}   \\   %	
    				Decentralized		    &\ding{51}      &\ding{51}     &\ding{51}    &\ding{55}        &\ding{51}	 	\\
    				\Xhline{1.0pt}
    			\end{tabular}
    			\caption{Comparison of contact tracing approaches}\label{tab:approaches}}
    	\end{floatrow}
    \end{table}
    A comparison of our approach with prior work is outlined in Table \ref{tab:approaches}. \texttt{CoAvoid} is an edge-based contact tracing approach based on GAEN. Experiments have proved that \texttt{CoAvoid} can protect the privacy of patients while resisting attacks. Meanwhile, as the data users need to upload and download has been significantly reduced, \texttt{CoAvoid} can achieve a rapid velocity in data transmission.
\section{Operation \& Threat Model}
\label{sec:motivation}
In this section, we illustrate how existing GAEN-based contact tracing applications operate by providing a COVID-19 running example.
Besides, by demonstrating the threat models,
we point out that these applications have potential privacy and security risks due to misuses of GAEN API.

\subsection{Running Example}
\label{sec:motivation_example}
To illustrate how existing GAEN-based contact tracing applications operate, we describe an running example in the COVID-19 pandemic.

GAEN API utilizes BLE technology on mobile devices to conduct contact tracings. The mobile device will randomly generate a Daily Tracking Key (DTK) during the operations, a unique device identifier that will change every 24 hours.
For a particular period $i$ of a day, the device uses some common deterministic function $f$ based on the DTK to derive the Rotating Proximity Identifiers (RPIs), \textit{privacy-preserving} identifiers that are sent in Bluetooth Advertisements, which will be replaced every 10 to 20 minutes. In other words $RPI=f(DTK,i)$. The device only needs to store the DTK with this operation because the actual temporary proximity identifier can always be reconstructed from DTK. A device with GAEN-based contact tracing enabled will broadcast RPIs to the surrounding users from time to time. Simultaneously, it collects and stores other devices' RPIs locally in a list $L$. Generation, propagation, and collection of RPIs occur automatically at the operating system level. Still, they are only allowed if the user notifies the application by installing exposure and setting the necessary permissions. Apple and Google don't allow exposure notification apps to access specific locations on users' devices. By default, public notification is disabled on both IOS and Android. When enabled, the database of DTKs and the received identifiers are stored in the operating system layer, ensuring that any applications the user installs cannot access the data directly.

\begin{figure}[htp]
	\centering
	\includegraphics[width=3.2in]{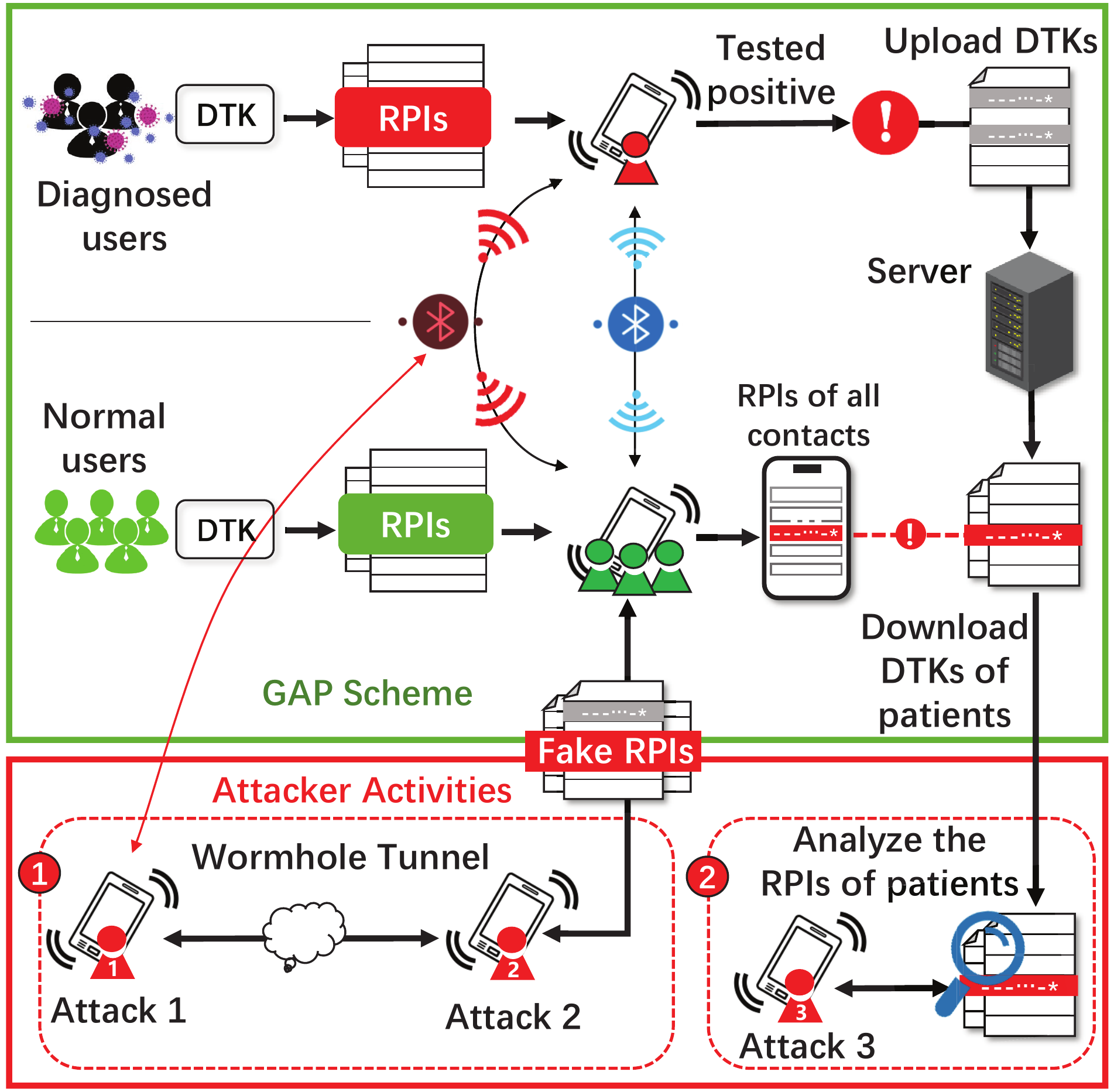}
	\caption{How GAEN-based contact tracing applications operates and two security \& privacy threats towards these schemes.}
	\captionsetup{font={scriptsize}}
	\label{fig:scenario}
\end{figure}

If a user has tested positive for COVID-19, GAEN-based contact tracing will upload its DTK to the central server with the user's permission. A central server will aggregate DTKs of confirmed COVID-19 patients and store such information in a list $L_{DTK}$ openly for 14 days. Other users can easily download $L_{DTK}$ and reconstruct the corresponding RPI list $L_{p}$.
By comparing list $L_{p}$ and list $L$, GAEN-based contact tracing applications can calculate a risk score based on many different factors (\eg contact time, distance between the two devices) to quantify the possibility of COVID-19 infection. If the risk score breaches a threshold, the contact tracing application will raise a warning for COVID-19 exposure.

Although this operation model is effective for contact tracing, it has several potential vulnerabilities:
(1) When the user downloads the DTKs of the patient on the server for risk assessment, the geographical location cannot be used to verify whether the user has been in contact with the patient. That is, the authenticity of the information cannot be verified, and thus wormhole attacks may occur;
(2) The server stores DTKs of patients openly, leading to invasions of patients' privacy;
(3) This running example assumes all users are benign, without any security designs to tackle malicious operations.
We elaborate on the threat models in Section~\ref{sec:2.2} to demonstrate how attackers can exploit these vulnerabilities.

\subsection{Threat Model} 
\label{sec:2.2}
\info{
    \item \textit{User forgery}. Attackers can legally obtain user identities to carry out attacks. Furthermore, it is difficult for normal users to identify the attacks and defend against them when exchanging information.
}
In our system, we assume that the central servers are considered as semi-trusted, in other words, \emph{honest but curious}. More specifically, these servers execute specific protocols but are curious about the content of user data. Moreover, all users’ devices can be divided into two classes: one is trustworthy devices and the other is malicious devices used to attack the naive devices of the users. Thus, according to the information view to the malicious devices, i.e., adversaries, we consider two types of attacks as following:
\begin{itemize}
	\item \textit{Wormhole attack}. A wormhole attack is a particular type of relay attack. An adversary $A$ can pretend to be naive devices and actively collect RPIs from the surrounding environment of one physical location and send them to the other end of the tunnel, namely, A cooperative adversary $B$ at another physical location. Then, $B$ propagates this information locally, misleading nearby devices as if two physical areas were merged into one logical area, making two groups of unrelated users seem to have touched. 
	
	In addition, if a directional gain antenna is used on the attacking device, the attacker can target specific users, showing them in different physical locations across the network. For example, an attacker can transmit a large number of false identifiers to users so that a specific user can collect a large amount of contact information and have many false contacts, which ultimately leads to a high probability that the user will be identified by the system as a high-risk user and may be self-isolated due to warnings. In addition, victims may undergo unnecessary medical tests due to the warning, which puts testing pressure on the capacity of medical centers. Using a wormhole device again, an attacker can create a significant virtual spread terror attack. While this has no direct impact on individual users, it is important for social stability. It also strains existing testing capacity and distracts government health workers from actual cases running in parallel.
	\item \textit{Privacy analysis attack}. Patients can be analyzed and identified, which is a risk for those who wish to remain anonymous. Because of the specific purpose of contact tracing applications, it is impossible for any application to avoid distance-based identification, even in the previous manual tracing of patient infection chains, especially now that GAEN exposes patient information on backend servers that may identify patients. For example, a malicious party can receive broadcasts simultaneously in multiple different locations, including those that the positive patients visited. Using this information and the DTKs of all recently submitted diagnostic users publicly available on a central server, malicious users can analyze the historical times and locations of active citizens to obtain information such as a patient's workplace, home address, and identity. In addition, since the DTK changes every 24 hours, it does not seem possible to track users for long periods of time. However, because patients need to upload DTKS for 14 days and some users have typical travel patterns, there are obvious similarities even on different days. For example, a user with a stable job has a fixed time and travel route from home to work every day, so it is possible to link and track some patients in this situation. This will reveal more personal information and activities of the target user and provide ample opportunity to use the additional public information that may be available to remove the anonymity of the user concerned.  In addition, de-anonymization is made easier if an adversary has access to additional information about a user's social relationships, such as the social graph of an online social network. Moreover, the semi-trusted central servers also have chances to monitor and analyze diagnosed  data of users for privacy mining.
\end{itemize}

\iffalse
\subsection{Security \& Privacy Requirements} \label{2.3}
To develop a privacy-preserving and secure contact tracing approach under the two types of attacks mentioned above, our scheme should satisfy the following secure and privacy requirements:
\begin{itemize}
	\item \textit{Location privacy}. Locations of users should not be obtained directly on condition that the devices are lost.
	\item \textit{Location verification}. Location information received from nearby should be verified to ensure it’s not forwarded.
	\item \textit{User privacy}. Users’ privacy should not be leaked or analyzed from information publicly stored by any malicious devices or by semi-trusted central servers.
\end{itemize}
\fi

\section{System Design} 
\label{sec:design}

To protect patients' privacy (including their personal information and social relationships) and to resist attacks that create false contact information to disrupt the system mechanisms, we proposes an edge-based, practical, and privacy-preserved system for contact tracing.

\subsection{Approach Overview} 
\label{3.1}

\begin{figure}[htp]
	\centering
	\includegraphics[width=0.95\textwidth]{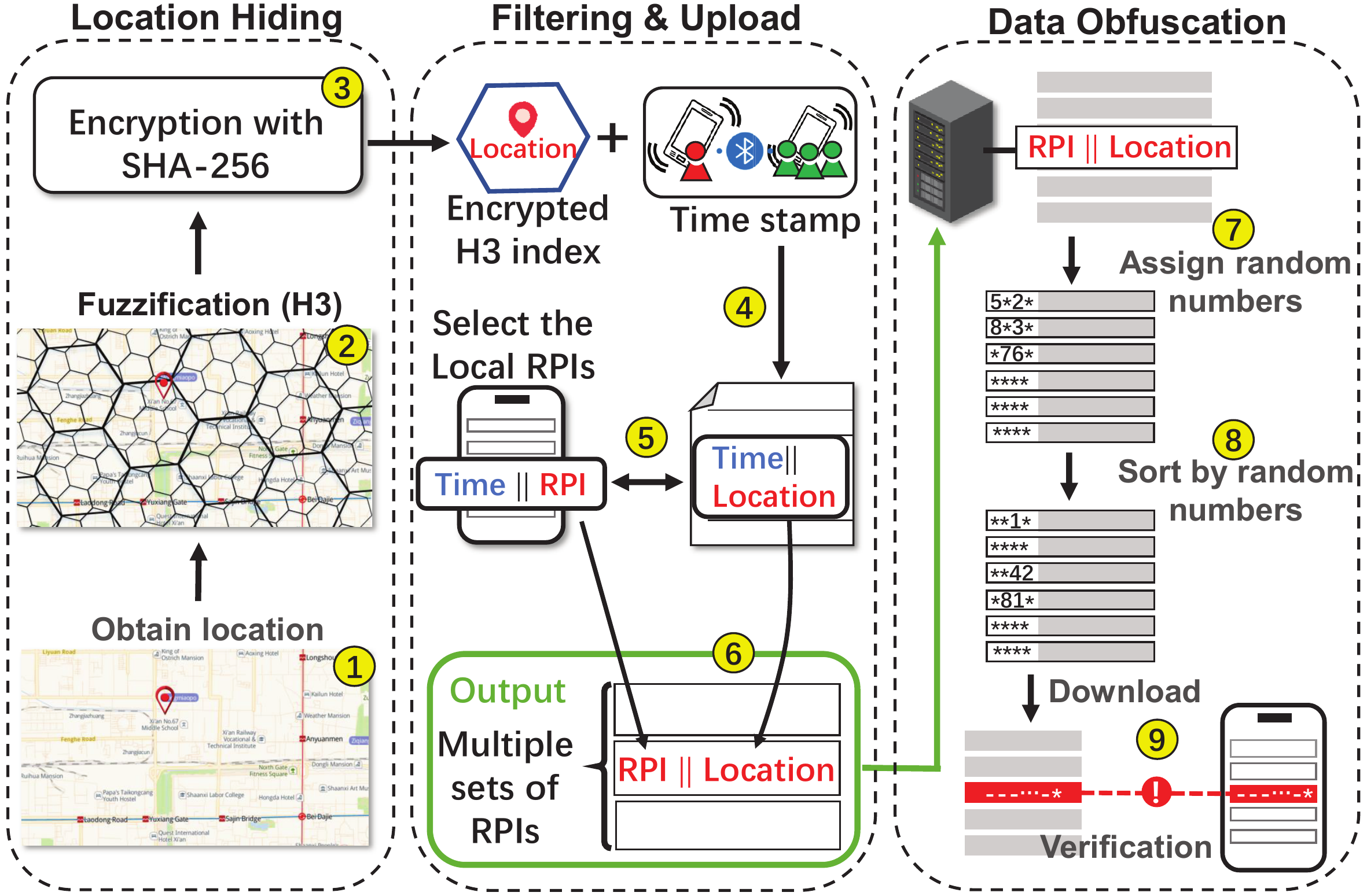}
	\caption{CoAvoid system overview.}
	\captionsetup{font={scriptsize}}
	\label{fig:approach}
\end{figure}

\texttt{CoAvoid} is an edge-based contact tracing system based on GAEN. It can efficiently protect the privacy of users and patients. \texttt{CoAvoid} can also resist wormhole attacks and achieve high accuracy on dynamic user trajectory matching with low system and time overhead, which fixes the deficiency of most GAEN-based contact tracing applications. Figure \ref{fig:approach} illustrates the framework of \texttt{CoAvoid}. Moreover, the tracing procedure of our system mainly consists of three main steps: \emph{Location Hiding}, \emph{Data Filtering \& Upload}, and \emph{Data Obfuscation \& validation}.

First, our approach anonymously collects the location data and pre-processes it to hide private information. Users save the timestamps, RPIs received from others, and their geographic locations in local devices when they generate a DTK and frequently send RPIs to surrounding users. To ensure the safety of patient location information, we obtain the longitude and latitude information through GPS (\emph{Step 1}) and leverage the H3 geospatial indexing system \footnote{https://h3geo.org/docs} to fuzz the location (\emph{Step 2}). Next, to make reverse operations that cannot obtain the original geographic locations of the users, we use $f_{SHA256}$ to hash the previous step (\emph{Step 3}).

Moreover, to reduce the exposure of positive COVID-19 cases information, \texttt{CoAvoid} only uploads the selected RPIs. When a user has tested positive for COVID-19, \texttt{CoAvoid} filters out the timestamps and hashed H3 location index of contact information between the patients and other users (\emph{Step 4}). After comparing the timestamps between the broadcasting and contact information, \texttt{CoAvoid} selects the RPIs in the broadcasting information (\emph{Step 5}) and combines them with location to upload the new joint information to the edge server (\emph{Step 6}).

Finally, the edge server will obfuscate the stored information to reduce the possibility of attackers analyzing the patients' data. The server will then distribute a random serial number to all the data uploaded by the patient before other users download the information (\emph{Step 7}). Furthermore, the server reorders the serial number and stores the data to complete the obfuscation (\emph{Step 8}). Other users download the obfuscated data for verification, ensuring the privacy of patients is not leaked while accurately matching users' locations(\emph{Step 9}).

In this section, we elaborate \texttt{CoAvoid} system according to the main three steps and describe the inner relationships. We also give the algorithms of our approach to explain it in detail. 

\subsection{Location Hiding}
The DTKs are generated every day on the user's device, and the RPIs to be broadcast during the day are calculated based on DTKs. Each DTK is generated independently by the cipher random number generator, and the RPI needs to be updated every 15 minutes. Specific operations such as encryption technology can refer to official documents \footnote{https://covid19-static.cdn-apple.com/applications/covid19}. The device will periodically broadcast RPIs to surrounding users every two minutes according to the GAEN scheme. Surrounding users will receive RPIs and save the timestamps, received RPIs, and location information locally for subsequent verification. Besides, users also need to save Bluetooth signal strength data for further risk rate calculation.

For privacy reasons, users will not use GPS coordinate $l$ ($l=(x,y)$) directly for verification. The location information $l'$ uploaded and verified by patients was obtained by the Equation \eqref{equation1}, where $f_{H3}$ represents a Geospatial division system, and $f_{SHA256}$~\cite{barker2002secure} refers to hash using the SHA-256 hash algorithm. 

\begin{equation}\label{equation1}
\begin{split}
f_{H3}(l) & =H_{index}(\{B|(x,y)\in B \land B \subsetneq L_{H3} \}), \\
l' & = f_{SHA256}(f_{H3}(l)).
\end{split}
\end{equation}

\begin{figure}[htp]
	\centering
	\includegraphics[width=3.4in]{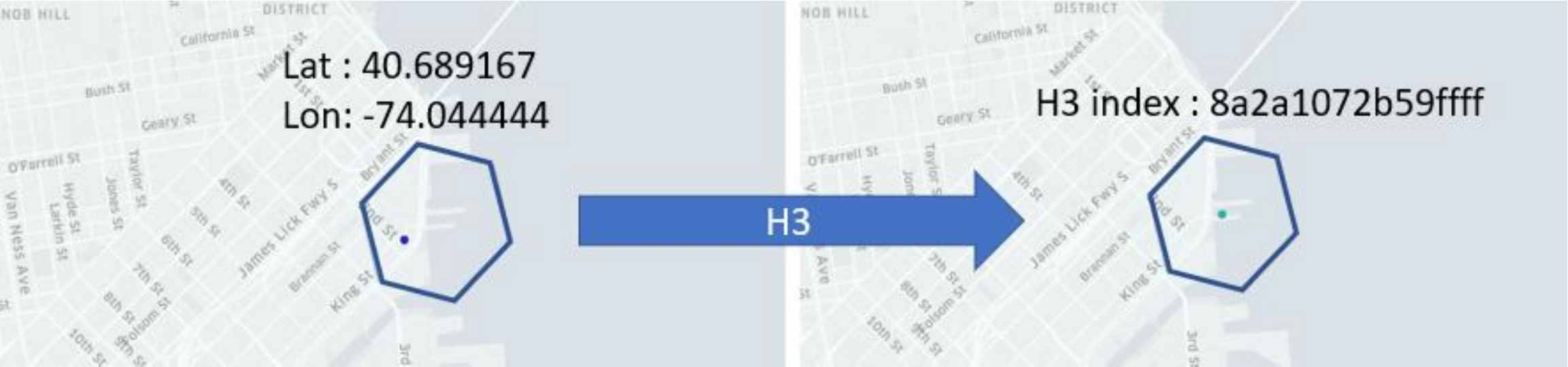}
	\caption{H3 index conversion diagram.}
	\captionsetup{font={scriptsize}}
	\label{fig:figure3}
\end{figure}

To fuzz the geographical positions of users and to solve the problem of inaccurate GPS location, we use Uber H3 ($f_{H3}$) to fuzz the geographical position. 
The H3 geospatial scale system is a multi-precision hexagonal spherical scale system with hierarchical linear indexing, whose core library provides functions for converting between coordinates and H3 geospatial indexes. 
The H3 system assigns a unique hierarchical index to each unit. Each level corresponds to a different resolution, with 16 resolutions of 0-15. The higher the resolution, the smaller the range corresponding to each H3 index, which means the range of hexagon region division is more accurate.
As is shown in Figure \ref{fig:figure3}, at a specific resolution, the geographic location information will have a corresponding H3 index. 
Under the high resolution, different GPS information under the H3 system corresponding index area will be similar but not corresponding. However, under the appropriate H3 resolution, two neighboring users can be calculated by the geospatial scaling system and generate the same hexagon index, which can be used for matching.
When patients upload information, we use H3 index to replace GPS information of users to blur location information. This can normalize the coordinates of adjacent GPS points so that the factual position information can be hidden.

Although coarse-grained GPS coordinates can already prevent users' specific coordinates from being leaked or analyzed, the risk of reversing conversion and abuse still remains.
Therefore, to prevent the privacy problem caused by information disclosure, we use the SHA-256 hash function to hash the generated H3 index. Due to the use of a one-way hash function for encryption cannot be easily reversed to expose the user's exact location information. In this way, the user can match the hashed data directly to the patient data on the server to verify whether the two users are close or have had contact.

\begin{algorithm}
	\caption{Location Hiding}
	\LinesNumbered
	\KwIn{Latitude $lat$ and longitude $lon$ in GPS information}
	\KwOut{Encrypted location information $L_{hide}$}
	$L_{h3}$ = $GeoCoord (lat, lon)$\;
	$L_{h3Index}$ = $geoToH3 (L_{h3}, resolution)$\;
	$L_{hide}$ = $SHA$-$256 (L_{h3Index})$\;
	\Return{$L_{hide}$}
	\label{alg:hiding}
\end{algorithm}

The complete process is shown in Algorithm \ref{alg:hiding}. The timestamp and geographic location information are saved when the user receives the surrounding RPI information while the system runs. Before uploading patient information, the device obtains the longitude and latitude coordinates of the user. After that, it obtains the geospatial index at a suitable resolution based on these GPS coordinates, and finally, the index is hashed using the SHA-256 function, and the final uploaded geographic location information is the hashed data.

\subsection{Data Filtering and Upload}
When a user has tested positive for COVID-19, the device needs to upload information from the last 14 days to a edge server after the user's authorization. The information must be filtered to prevent privacy from leaking. 
The main idea of the filtering approach is that it must upload the RPIs of patients who have contact with other users. 
There are three main steps of the filtering scheme, which is illustrated in Figure \ref{fig:figure4}: 

\begin{figure}[htp]
	\centering
	\includegraphics[width=2.9in]{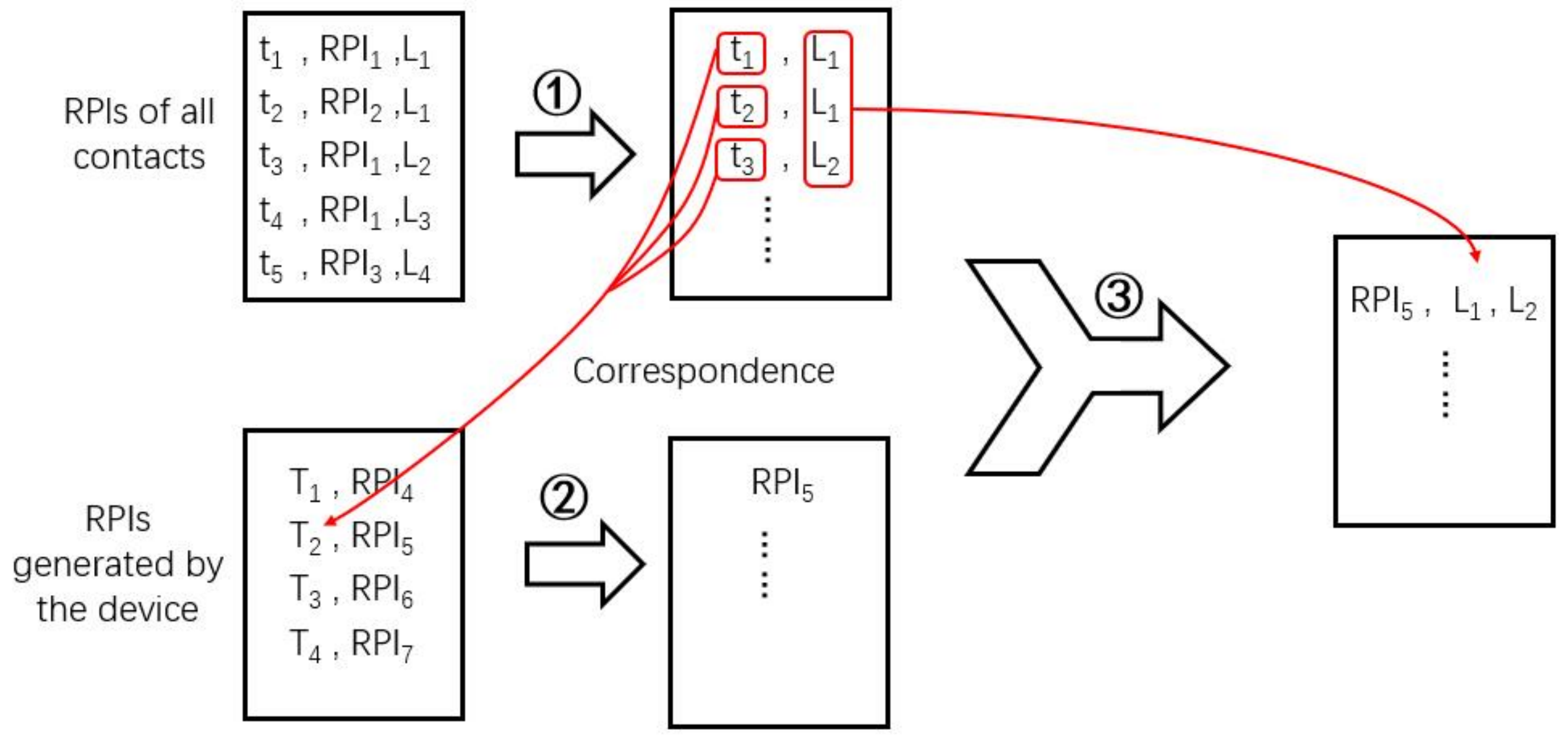}
	\caption{Data filtering and upload.}
	\captionsetup{font={scriptsize}}
	\label{fig:figure4}
\end{figure}

\begin{itemize}
	\item \textit{Step 1}: 
	We need to screen out the RPIs that have been exchanged and stored in the patient's device. If the same RPI information has been recorded one time, the user corresponding to the RPI is identified as having potential infection risk. Then, the time stamps $t_i$ and location information $L_i$ corresponding to the RPI is extracted. The location at this time is the information hashed by SHA-256.
	\item \textit{Step 2}: 
    Our approach compares these time stamps $t_i$ with the RPI generated by the device of the COVID-19 patient. Then, we need to determine the time period $T_i$ corresponding to these timestamps $t_i$ and record the RPI of the patient corresponding to this period. 
	\item \textit{Step 3}: 
	We integrate the location information $L_i$ corresponding to the time stamp $t_i$ in \emph{Step 1} with the RPI of the patient selected in \emph{Step 2} to form the new RPI, which will be uploaded to the server.
\end{itemize}

\begin{algorithm}
	\caption{RPI Filtering and Upload}
	\LinesNumbered
	\KwIn{The RPIs $ER=\{er_1,er_2,...,er_i\}$ generated by the patient and other users exchanged,The RPIs $GR=\{gr_1,gr_2,...,gr_i\}$ generated by patient equipment, $gr_i=\{Time||RPI\}$}
	\KwOut{Reorganized RPI information $RR=\{rr_1,rr_2,...,rr_i\}$}

	\For{each $er_i$ $\in$ ER}
	{
		\If{The records of $er_i$ $\geq$ 1}
		{
			Extract the timestamp $t_i$ and location $L_i$ in $er_i$\;
		}
	}
	
	\For{$T$ = each $gr_i.Time()$}
	{
		\For{each $t_i$ $\in$ t}
		{
			\If{$t_i$ $\in$ $T$}
			{
				Extract the $RPI_i$ in $gr_i$\;  
			}
		}
	}
	\For{each $t_i$ $\in$ $t$ and each $L_j$ $\in$ $L$}
	{
		\If{i = j}
		{
			$rr_i$ = [The RPI $RPI_k$ corresponds to the timestamp $t_i$ $||$ $L_j$]\;	
		}
	}
	\Return{$RR$}
	\label{alg:filtering}
\end{algorithm}

Compared with directly uploading DTK in the GAEN scheme, our approach can reduce patient information leakage and privacy analysis risk. In other GAEN-based applications, anyone can use DTK to calculate all the RPI used by a patient in a day. That is, the patient RPI is public data. If the attacker has collected enough RPI information in advance at different locations, he can compare the patient's RPI with the RPI information collected to get the time and location of the patient's encounter with the attacking device. Therefore, he could figure out the patient's residence, workplace, occupation, and social relations.
So attackers can use DTK to de-anonymize patients.

Our approach screens out RPIs of patients who have long periods of contact with other users to upload. First of all, it reduces the number of RPI uploads by patients, i.e., it fundamentally reduces the amount of data disclosed by patients and makes the RPI information disclosed by patients discontinuous in time by this method. Filtering RPI reduces the risk of attackers tracking patient contact information through the timeline and compensates for the disadvantages of direct DTK uploads. For example, suppose an attacker collects RPI information at multiple locations, and the patient visits these locations in a single day. In this case, after multiple days, the attacker can identify the patient by matching the RPI computed with the patient's publicly available DTK, but when the patient uploads a filtered discrete RPI, the attacker cannot assume that these matched RPI's belong to the same patient, thus reducing the risk of identifying the patient. Secondly, the RPI information of all patients will be confused in the edge server. Once the RPI information of multiple people is mixed, the adjacent RPI stored will not belong to the same patient. Even if the attacker obtains the discrete point of the time and place of the patient group, it cannot connect the historical action trajectory of a single patient and obtain the social circle of a single patient. It protects the patient's privacy and achieves the purpose of preventing the attacker from anonymizing the patient. Algorithm \ref{alg:filtering} describes the filtering approach to the RPIs of positive patients.

\subsection{Obfuscation and Verification}
The RPIs of all patients will be uploaded to the edge server. To prevent attackers from analyzing all public data, the edge server obfuscates all the received data by disrupting the default upload order stored in the server. This makes the adjacent stored data not the same as the information uploaded by the same patient.

Algorithm \ref{alg:server} shows the data obfuscation method. When the patient uploads RPIs, the edge server assigns a random sequence number $S_i$ for each message. Before other users download all the data, the edge server reorders the RPIs according to $S_i$, which completes the information shuffle. This process will only be executed on the server side.
After data obfuscation, the random serial number $S_i$ will be deleted. Thus, the user application will only get the RPI and geographic location information.

\begin{algorithm}
	\caption{RPI Obfuscation on Server}
	\LinesNumbered
	\KwIn{The RPIs $RR=\{rr_1,rr_2,...,rr_i\}$ uploaded after screening and reorganization}
	\KwOut{Obfuscated RPI information $CR=\{cr_1,cr_2,...,cr_i\}$}
	
	\For{each $rr_i$ $\in$ RR}
	{
		Generate a random number $S$\;
		$rr_i$ = [$S$ $||$ $rr_i$]\;
	}
	
	$CR$ = $RR$ sort by $S$\;
	
	\Return{$CR$}
	\label{alg:server}
\end{algorithm}

The user's apps will periodically download data from the edge server for information verification. Once matching the same RPI and geographic information, this indicates that the user has likely been exposed to confirmed COVID-19 cases.

\begin{figure}[htp]
	\centering
	\includegraphics[width=2.0in]{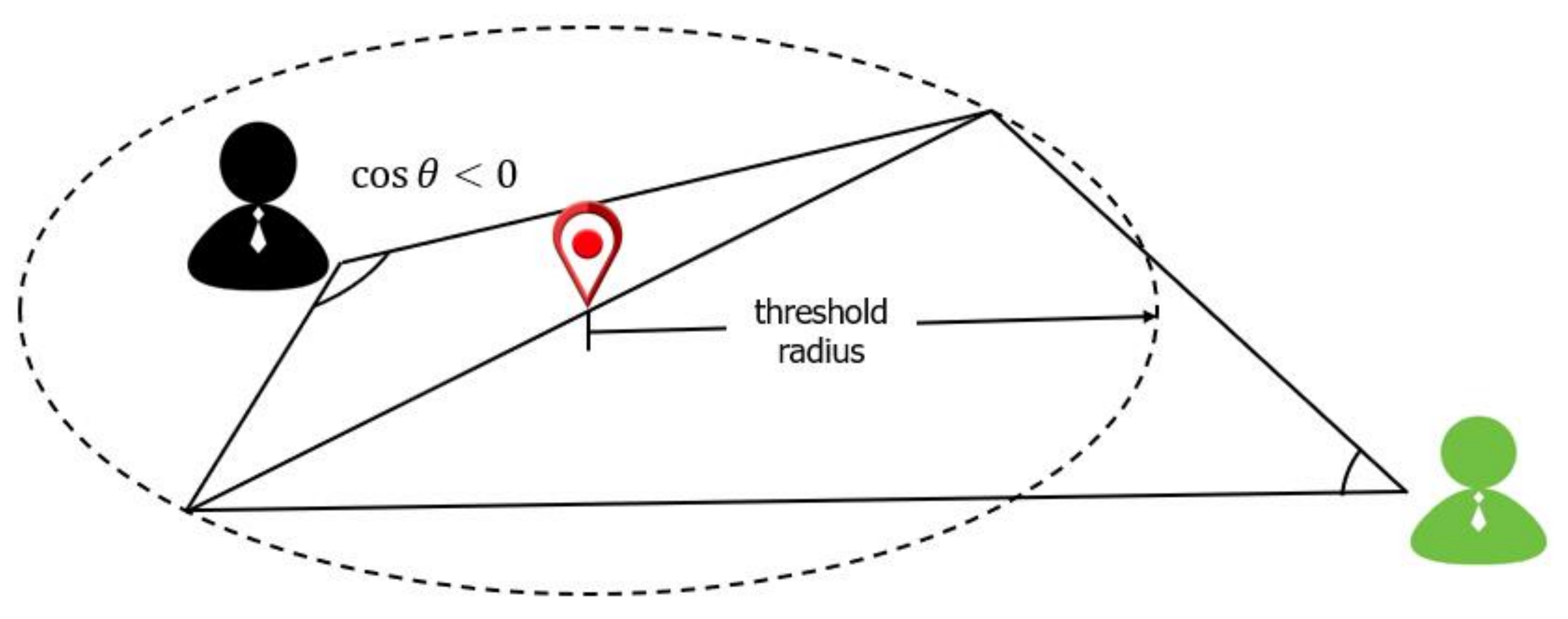}
	\caption{Lightweight cryptographic matching method.}
	\captionsetup{font={scriptsize}}
	\label{fig:lightweight}
\end{figure}

A more precise lightweight cryptographic matching is later performed to determine the user's location with respect to the patient, it is necessary to determine whether the user is within the threshold radius, as shown in Figure \ref{fig:lightweight}. This requires the patient and user to push cryptographic parameters to each other, calculate the vector point product consisting of the user and patient security circle diameter endpoints, and then determine whether the user is in the same physical area by the cosine value positive or negative. 

\begin{itemize}

  \item [1)] 
Before performing the algorithm for each step, CoAvoid makes the following preparations. The first step is to generate bilinear pairing parameters $q,g,G,G_T,e$ by choosing bilinear mapping $e:G×G=G_T$, where G is a group of order $q$, $q$ is a large prime, and $g$ is the generator of $G$. The patient then generates the private key $SK_s\in\mathbb{Z}_q^\ast$, the public key $PK_s=g^{SK_s}$, and security parameters $k_1,k_2,k_3,k_4$. The selection of parameters needs to satisfy Equation \eqref{k}'s correctness condition so that data integrity can be maintained before and after the data forwarding (\eg $k_1=800$, $k_2=300$, $k_3=128$ and $k_4=128$). The patient generates large prime numbers $\alpha \in \{0,1\}^{k_2}, p\in \{0,1\}^{k_1}$ and random numbers $s_m\in\mathbb{Z}_q^\ast, a_{m_j}\in \{0,1\}^{k_3}$. The user generates random numbers $r_{k||m}\in\{0,1\}^{k_4}$, the private key $SK_p\in\mathbb{Z}_q^\ast$, and the public key $PK_p=g^{SK_p}$. Finally, the patient and the user negotiate the session key.
  \begin{equation}\label{k}
\begin{split}
\revision{k_4+max((2 \cdot k_2+44+2),(k_2+22+k_3+2)) < k_1}\\
\revision{k_4+max((2 \cdot k_2+44+2),(k_2+44+1+k_3)) < k_1}\\
\revision{k_4+k_3+44+2 < k_2}
\end{split}
\end{equation}

   \item [2)] 
    From the patient's true location coordinates $P_{IP_m}(x_{IP_{m_0}},y_{IP_{m_0}})$, two points $P_{IP_{m_1}}(x_{IP_{m_1}},y_{IP_{m_1}})$, $P_{IP_{m_2}}(x_{IP_{m_2}},y_{IP_{m_2}})$ are generated according to the distance threshold, satisfying the following conditions. 
\begin{equation}\label{equation19}
\begin{split}
2 \cdot x_{IP_{m_0}} = x_{IP_{m_1}} + x_{IP_{m_2}}\\
2 \cdot y_{IP_{m_0}} = y_{IP_{m_1}} + y_{IP_{m_2}}
\end{split}
\end{equation}
  \item [3)]The patient device generates a random number $s_m$, $a_{m_j}$ and calculates the encrypted information. $a_{m_j}$ should be replaced after each generated $EN_m$, to ensure that $a_{m_j}$ is not repeated.
\begin{equation}\label{equation4}
\begin{split}
EN_m = EN_{m_1} \lVert EN_{m_2} \lVert EN_{m_3} \lVert EN_{m_4}\\ \lVert EN_{m_5} \lVert EN_{m_6} \lVert EN_{m_7}
\end{split}
\end{equation}

\begin{equation}\label{equation5}
\begin{split}
EN_{m_1} = s_m \cdot (x_{IP_{m_1}} \cdot \alpha + a_{m_1})\bmod p \\
EN_{m_2} = s_m \cdot (y_{IP_{m_1}} \cdot \alpha + a_{m_2})\bmod p \\
EN_{m_3} = s_m \cdot (x_{IP_{m_2}} \cdot \alpha + a_{m_3})\bmod p \\
EN_{m_4} = s_m \cdot (y_{IP_{m_2}} \cdot \alpha + a_{m_4})\bmod p \\
EN_{m_5} = s_m \cdot (x_{IP_{m_1}} \cdot x_{IP_{m_2}} \cdot \alpha + a_{m_5})\bmod p \\
EN_{m_6} = s_m \cdot (y_{IP_{m_1}} \cdot y_{IP_{m_2}} \cdot \alpha + a_{m_6})\bmod p \\
EN_{m_7} = s_m \cdot (\alpha + a_{m_7})\bmod p 
\end{split}
\end{equation}
  
  \item [4)]The patient device is signed using Equation \eqref{equation6} and encrypted using Equation \eqref{equation7}, after which the server pushes the information to all users who have undergone coarse-grained screening. The user receives the message and obtains $p$, $\alpha$, $EN_m$ by decrypting the message and verifies that the data satisfies Equation \eqref{equation8}. Where $H()$ is the hash function, $E()$ is a secure encryption algorithm such as SM4, $SK_p$, $PK_p$ is the public-private key pair used for encryption and decryption, and $SID$ is the session key ID number.
 \begin{equation}\label{equation6}
\begin{split}
Sig_p = H(p \lVert \alpha \lVert EN_m \lVert timestamp \lVert SID)^{SK_p}
\end{split}
\end{equation}

 \begin{equation}\label{equation7}
\begin{split}
message = E(p \lVert \alpha \lVert EN_m \lVert timestamp \lVert SID \lVert Sig_p)
\end{split}
\end{equation}
 
  \begin{equation}\label{equation8}
\begin{split}
e(g,Sig_p) = e(PK_p, H(p \lVert \alpha \lVert EN_m \lVert timestamp \lVert SID))
\end{split}
\end{equation}
To ensure that the user's location information is not fraudulently obtained after the server is compromised by a malicious attacker, it is necessary to verify the encrypted information received. The patient location encrypted information that needs to be obtained satisfies Equation \eqref{equation9}. 

  \begin{equation}\label{equation9}
\begin{split}
EN_{m_1} + EN_{m_3} \neq 0\\
EN_{m_2} + EN_{m_4} \neq 0\\
EN_{m_7} \neq 0
\end{split}
\end{equation}

  \item [5)]The user obtains each value of $EN_{m_j}$ through $EN_m$, generates a random number $r_{k \lVert m}$, and calculates the user location encryption information $A_{k \lVert m} = A_{k \lVert m_{1}} \lVert A_{k \lVert m_{2}}$. Where $(x_u, y_u)$ is the GPS coordinate point of the user.
    \begin{equation}\label{equation10}
\begin{split} 
A_{k\lVert m_{1}}=r_{k\lVert m}\cdot\alpha\cdot(x_u\cdot(EN_{m_1}+EN_{m_3})\\+y_u\cdot(EN_{m_2} + EN_{m_4}))\bmod p\\%
A_{k\lVert m_{2}}=r_{k\lVert m}\cdot\alpha\cdot(EN_{m_5}+EN_{m_6}(x_u^{2}+\\y_u^{2})\cdot EN_{m_7})\bmod p
\end{split}
\end{equation}
 
 \item [6)]The user sends this information to the patient, and the patient device calculates the location relationship information $C_{k \lVert m}=C_{k \lVert m_{1}}-C_{k \lVert m_{2}}$. The calculation process is as follows.
     \begin{equation}\label{equation11}
\begin{split} 
B_{k \lVert m_{1}} = s_m^{-1}\cdot A_{k\lVert m_{1}} \bmod p\\
B_{k \lVert m_{1}} = s_m^{-1}\cdot A_{k\lVert m_{2}} \bmod p
\end{split}
\end{equation}

      \begin{equation}\label{equation12}
\begin{split} 
C_{k \lVert m_{1}} = \frac{B_{k \lVert m_{1}} - (B_{k \lVert m_{1}}\bmod \alpha^{2})}{\alpha^{2}}\\
C_{k \lVert m_{2}} = \frac{B_{k \lVert m_{2}} - (B_{k \lVert m_{2}}\bmod \alpha^{2})}{\alpha^{2}}
\end{split}
\end{equation}
 
 \item [7)]Determine if $C_{k \lVert m} < 0$ holds and if so, the user has had contact with the patient.
 
\end{itemize}

Information verification can eliminate the influence of wormhole attacks and relay attacks. RPI information is the only information that an attacker can touch and change. If an attacker changes the patient's RPI and broadcasts it to the surrounding devices in another place, the attacker creates a non-existent RPI. It cannot cause serious consequences, and the relay attack is invalid. If the attacker happens to modify the RPI of a patient, the attack can be resisted according to the wormhole attack. Furthermore, the user will verify information if the attacker transmits the RPIs from location $L_A$ to $L_B$. Although the same RPI is detected, the geographic location information does not match. Finally, the verification will not pass, and the wormhole attack can be invalid.

When a user has tested positive for COVID-19, the approach can use the Bluetooth strength data to calculate the exposure risk value. Using parameters to calculate the exposure risk value is to narrow the range of contacts who may be infected and to increase the accuracy of the application in identifying potential contacts to avoid the unnecessary panic of risk-free users.
If a user passes by the patient, the application will not warn it. If the risk is high, the system will notify the user with further instructions.
The calculation of the exposure risk value refers to the GAEN design \footnote{https://developer.apple.com/documentation/exposurenotification}. There are four parameters involved in calculating the risk score, which is shown in Equation \eqref{equation2}.

\begin{equation}\label{equation2}
\begin{split}
RiskScore = TRV \cdot DuRV \cdot DaRV \cdot ARV
\end{split}
\end{equation}

\begin{itemize}
	\item \textit{$Transmission\_Risk\_Value$ ($TRV$)}:
	It reflects the patient's condition and his impact on transmission risk. This value depends on the patient's symptoms, the time these symptoms first appeared, the level of disease diagnosis, or other judgments of the authority.
	\item \textit{$Duration\_Risk\_Value$ ($DuRV$)}: 
	Cumulative contact time between the user and the confirmed COVID-19 case.
	\item \textit{$Days\_Risk\_Value$ ($DaRV$)}: 
	The number of days since the last contact with the confirmed COVID-19 case.
	\item \textit{$Attenuation\_Risk\_Value$ ($ARV$)}: 
	Signal strength changes during contact to the confirmed COVID-19 case. Therefore, when the attenuation value is greater than 0, the weighted calculation will be performed according to the duration of each risk level, and the average of the total duration will be taken.
\end{itemize}

Each parameter is divided into nine levels: 0-8, which is also the value assigned to each parameter. Finally, the numbers of four parameters are multiplied to calculate the corresponding risk score of the user. 

\section{Analysis \& Evaluation}
\label{sec:analysis}

\subsection{Security \& Privacy Analysis}
In this section, we theoretically analyze \texttt{CoAvoid} from the perspectives of security and privacy.

\subsubsection{Attack Resistance}

Here, we prove that the fine-grained matching algorithm is secure. More specifically, with the existence of adversary $A$ and hypothetical challengers $P_c$ and $U_{cU}$, we prove the semantic security of the fine-grained matching algorithm by playing a game between the adversaries and challenges. Let $\varepsilon$ denote the algorithm. $\varepsilon_{EP}$ is the patient's encrypted location data in the algorithm, and $\varepsilon_{EU}$ is the user's encrypted location information in the algorithm.

First, we need to verify the security of the patient location information. $P_c$ generates the required system parameters, and $A$ selects two pairs of location information, $P_{IP_{cP_0}}$ and $P_{IP_{cP_1}}$, to send to $P_c$. After $P_c$ receives the information, it randomly selects a bit $b\in \{0,1\}$ and randomly selects a non-repeating $a_j(j=1\sim7)$. Then, it encrypts the location information with key $s$ and parameter $\alpha$, and returns $EN_b$ to $A$, where $EN_b=EN_{b_1}||\dots||EN_{b_7}$. We can calculate $EN_{b_j}$ using Equation \eqref{a1}.

 \begin{equation}\label{a1}
\begin{split}
EN_{b_1}=s\cdot(x_{IP_{cP_0}}\cdot \alpha+a_1)\mod p\\
EN_{b_2}=s\cdot(y_{IP_{cP_0}}\cdot \alpha+a_2)\mod p\\
EN_{b_3}=s\cdot(x_{IP_{cP_1}}\cdot \alpha+a_3)\mod p\\
EN_{b_4}=s\cdot(y_{IP_{cP_1}}\cdot \alpha+a_4)\mod p\\
EN_{b_5}=s\cdot(x_{IP_{cP_0}}\cdot x_{IP_{cP_1}}\cdot \alpha+a_5)\mod p\\
EN_{b_6}=s\cdot(y_{IP_{cP_0}}\cdot y_{IP_{cP_1}}\cdot \alpha+a_6)\mod p\\
EN_{b_7}=s\cdot(\alpha+a_7)\mod p
\end{split}
\end{equation}

After $EN_b$ is returned to $A$, $A$ transmits a bit $b'\in\{0,1\}$ to $P_c$ and tries to infer which location message $P_c$ encrypts.

$s$ is a randomly generated secrecy parameter of at least $400(0.5k_1)$ bits by $P_c$, while $a_j$ is a non-repeating $128(k_3)$ bits random number and a unique $a_j$ is used for each location message encryption. The location information within any of the encrypted messages can be obtained by the following formula, for example, the location information $x_{IP_{cP_0}}$ can be obtained by $\frac{((s^{-1}\cdot EN_{b_1})-a_1)\mod p}{\alpha}$. However, as both $s$ and $a_j$ are generated by the secure random number generator and are immediately encrypted, there are two enormous unknowns for each message of $A$. In addition, since each message is encrypted with a unique $a_j$ at each time, $A$ is unable to distinguish the messages encrypted by $P_{IP_{cP_0}}$ or $P_{IP_{cP_1}}$ Assume the keyspace for location information is $\partial(key)$ and the keyspace for the longitude and latitude information is $\partial(P_{IP})$(i.e., $s,a_j\in\partial(key)$ and $P\in\partial (P_{IP})$). We can then derive Equation \eqref{a2}:

\begin{equation}\label{a2}
\begin{split}
 &\rho r_{P_{IP_{cP_0}},P_{IP_{cP_1}}}((s,a_j,P)=EN_l)\\&=\frac{\#s,a_j\in\partial(key),s\cdot t~EN_{bj}(s,a_j,P)=EN_l}{2^{428}}
 \\&=Constant
\end{split}
\end{equation}

At this time, $SSadv[A,\varepsilon]:=\lvert\rho r(b=b')-\frac{1}{2}\rvert$ can be ignored, which satisfies the semantic security.

Next, we need to verify the security of the security of location information for ordinary users. Let $A$ select two pairs of location information, $P_{IP_{cU_0}}(x_{IP_{cU_0}},y_{IP_{cU_0}})$ and $P_{IP_{cU_1}}(x_{IP_{cU_1}},y_{IP_{cU_1}})$, and send them to $U_{cU}$. $U_{cU}$ receives the information and randomly selects a bit $b\in \{0,1\}$ and a non-repeating $r_{cU||pb}$. To hide the location information, $U_{cU}$ multiplies the location information $P_{IP_{cU_b}}$, the received patient-encrypted information $EN_m$, and $r_{cU||pb}$. After that, $U_{cU}$ sends $A_{cU||cPb}=A_{cU||cPb1}||A_{cU||cPb2}$ to $A$. $A$ then returns a bit $b'\in\{0,1\}$ to $U_{cU}$ and tries to infer which location information $U_{cU}$ encrypted. We can calculate $A_{cU||cPb}$ using Equation \eqref{a3}.

 \begin{equation}\label{a3}
\begin{split}
 A_{cU||cPb1}=r_{cU||pb}\cdot\alpha\cdot(x_{IP_{cU_b}}\cdot(EN_{cUb_1}+EN_{cUb_3})+\\y_{IP_{cU_b}}\cdot(EN_{cUb_2}+EN_{cUb_4}))\mod p\\
 A_{cU||cPb2}=r_{cU||pb}\cdot\alpha\cdot(EN_{cUb_5}+EN_{cUb_7}+(x_{IP_{cU_b}}^2+\\y_{IP_{cU_b}}^2)\cdot EN_{cUb_7})\mod p\\
\end{split}
\end{equation}

In this assumption, since the random number generator is secure and the generated $r_{cU||pb}$ is a non-repeating random number, $A$ cannot distinguish between $r_{cU||p0}$ and $r_{cU||p1}$. In addition, the user's location information $x_{IP_{cU_b}},y_{IP_{cU_b}})$ is used to encrypt the message. $A$ will consider that the encrypted information $A_{cU||cPb1}$ and $A_{cU||cPb1}$ generated by the user are unary cubic polynomial and binary cubic polynomial. Thus, $A$ cannot distinguish between the encrypted messages $P_{IP_{cU_0}}$ or $P_{IP_{cU_1}}$, which means $SSadv[A,\varepsilon]:=\lvert\rho r(b=b')-\frac{1}{2}\rvert$ can be ignored. Therefore, semantic security is satisfied here.

Suppose there is a patient $P$, two users $U_1$ and $U_2$, and a hypothetical adversary $A$, where $A$ tries to obtain the patient's $RPI$ information and propagate it to other users in other geographical locations. $A$ randomly selects the received $RPI$ and forwards it.

Assume that $\varepsilon$ is the location verification algorithm, $\varepsilon_1$ is the coarse-grained location verification algorithm, and $\varepsilon_2$ is the fine-grained location verification algorithm. $\varepsilon_1$ has a location space of $\partial(f_{SHA256}(f_{H3}(Location)))$ and an $RPI$ message space of $\partial(RPI)$, where $\forall RPI_P,RPI_A\!\in\! \partial (RPI)$ and $\forall loc_P,loc_{U_1 }\in \partial(f_{SHA256}(f_{H3}(Location)))$.

Adversary $A$ actively collects the $RPI$ of surrounding users through the BLE device and forwards it to other geolocation locations. In our method, user $U_1$, after downloading the information about patient $P$ from the server, will first verify whether the received location and their own location are $RPI$ in a common vicinity with algorithm $\varepsilon_1$, as shown in Equation \eqref{a4}.

 \begin{equation}\label{a4}
\begin{split}
&(RPI_P\parallel loc_P)\oplus(RPI_A\parallel loc_{U_1})\\&= \begin{cases}
RPI_P\oplus RPI_A &if~loc_P=loc_{U_1}\\
(RPI_P\oplus RPI_A)\parallel (loc_P\oplus loc_{U_1}) &if~ loc_P \neq loc_{U_1}
\end{cases}
\end{split}
\end{equation}

Since $f_{SHA256}$ is a strong collision-resistant hash function, its security and unforgeability are hard to break. Besides, the probability that the hash values of two location messages will conflict is negligible. Due to these factors, the advantage shown in Equation \eqref{a5} is negligible.

 \begin{equation}\label{a5}
\begin{split}
&Adv^{SHA256} =\\ &Pra[f_{SHA256}(f_{H3}(loc_P))=f_{SHA256}(f_{H3}(loc_{U_1}))]
\end{split}
\end{equation}

Besides, when user $U_2$ performs fine-grained location verification V, attacker A cannot break our system due to the security of the lightweight matching algorithm. This means that, $SSadv[A,\varepsilon]:=\lvert \rho r(P_{IP_P}=P_U )-\frac{1}{2}\rvert$ can be ignored, where $P_{IP_P}$ is the coordinate of the patient and $P_U$ is the coordinate of the user. Therefore, our system can guarantee secure location verifications even in the presence of wormhole attacks.

\subsubsection{Privacy Analysis}
Our approach utilizes various methods to protect user privacy, such as processing all the data anonymously, using encryption algorithms to prevent data leakage, and using fuzzification and obfuscation to shelter sensitive data.

Assume that $\varepsilon_{RPI}$ is the encryption algorithm of $PRI$ with location space $\partial(f_{SHA256}(f_{H3}(Location)))$ and $RPI$ message space $\partial(RPI)$, where $\forall RPI_1,RPI_2\in \partial(RPI)$ and $\forall loc_1,loc_2\in \partial (f_{SHA256}(f_{H3}(Location)))$.

Before the user uploads the location information, the location information will be first fuzzed by Equation \eqref{equation1}, then the fuzzy range of the original location reaches ${\frac{3}{2}\sqrt{3}r}^2$, where $r$ is the side length of the hexagon when the processing is carried out.

From the previously mentioned loc encryption algorithm, we know that $(RPI_1\parallel loc_1)\oplus (RPI_2\parallel loc_2)$ can be used to determine whether the user and the patient are in the same hexagonal region. In worst case, attacker A can analyze and attack the system within the same hexagonal region to find out the $RPI$ information of the same user.
From the BLE information generation algorithm, we can derive the following equation:

 \begin{equation}
\begin{split}
RPIK_i=f_{SHA256}(DTK_i,NULL,UTF8("EN-RPIK")),
\end{split}
\end{equation}

where the identity key $DTK_l$ that updated every 15 minutes by each user is a secure pseudo-random number. $\partial(DTK)$ denotes the keyspace for $DTK_l$, where $\forall DTK_l \in \partial(DTK)$ and $\forall RPI_l \in \partial(RPI)$.

As Equation \eqref{a6} shows, $f_{SHA256}$ is a strong collision-resistant hash function, it is thus difficult for attackers to obtain the information sent by the user with brute force attacks.

\begin{footnotesize} 
\begin{equation}\label{a6}
\begin{aligned}
&\rho r_{DTK_{l}}\left(\left(DTK_{l},\right.\right.\left.UTF8(\mathrm{EN}\!-\!\mathrm{RPIK}))=RPIK_{ex}\right)\\
&=\!\frac{\!\#\! D\!T\!K_{l}\! \in \!\partial(\mathrm{DTK}), s\!\cdot \!t \!~\!f_{SHA256}\!\left(\!D\!T\!K_{l},U\!T\! F(\mathrm{EN}\!-\!\mathrm{R\!P\!I\!K\!})\right)\!=\!R\!P\!I\! K_{e\!x}}{|\mathrm{K}|} \\
&<\frac{1}{2^{128}}\!
\end{aligned}
\end{equation}
\end{footnotesize}

The generated $RPIK_l$ is a random string, which is very unpredictable. $RPIK$ is the key for the next round of encryption to obtain $RPI_l$, and its keyspace is noted as $\partial(RPIK)$.

\begin{equation}\label{a7}
\begin{aligned}
\text {PaddedData}_{l}\!=\!U T F 8(\mathrm{EN}\!-\!\mathrm{RPI})\!\parallel\! 0 x 000000000000\!\parallel\! E N I N_{j}
\end{aligned}
\end{equation}

In the above equation, $ENIN_j\!=\!ENIntervalNumber(j)$ is the serial number corresponding to each $RPI$, ranging from 1 to 96.
As stated above, $RPIK_l$ is unpredictable, and the broadcast message $PaddedData_l$ satisfies the AES security padding condition, as shown in Equation \eqref{a8}.

\begin{footnotesize} 
\begin{equation}\label{a8}
\begin{aligned}
&\rho r_{\mathrm{RPIK}_{l}, paddedData_{l}}\left(\left(\mathrm{RPIK}_{l}, \text{ paddedData}_{l}\right)=R P I_{e x}\right)\\
&=\frac{\# R P I K_{l} \!\in\! \partial(\mathrm{RPIK}), s \!\cdot\! t ~ f_{AES\!-\!128}\left(D T K_{l}, \text { paddedData}_{l}\right)\!=\!R P I_{e x}}{|\mathrm{K}|}\\
&=\operatorname{pr}\left( f_{AES\!-\!128}\left(\mathrm{RPIK}_{l},  paddedData_l\right)=RPI_{ex}\right)
\end{aligned}
\end{equation}
\end{footnotesize}

At this point, $\operatorname{SSadv}[A, \varepsilon]:=\left|\operatorname{pr}\left(\varepsilon\left(D T K_{l}\right)=R P I_{l}\right)-\frac{1}{2}\right|$ can be ignored, which satisfies the semantic security, i.e., attacker $A$ cannot tell which information belongs to the same patient. As a result, our algorithm can guarantee the security of patients' personal information.

\subsection{Evaluation}
In our experiment, we evaluate our system from the perspectives of privacy, security, and efficiency. We compared \texttt{CoAvoid} with the centralized solution TraceTogether~\cite{stevens2020tracetogether}, the distributed solution COVIDWISE, DP-3T~\cite{troncoso2020decentralized} and Immuni. COVIDWISE and Immuni were developed by the Virginia Department of Health and Italy based on GAEN API.

\subsubsection{Dataset and Configuration}

In our experiment, we simulated the daily interactions of thousands of users with a simplified random walk model. The random walk model is one of the most widely used mobility models for network behavior analysis. Therefore, we utilized a simplified variant of the random walk model in our setting. More specifically, we set up $n$ locations as user contact locations, and each location is assigned with an actual GPS geographic location. The distances between these geographic locations may vary but will generally remain within the confines of a city. According to the random walk model, each user is required to go to other places randomly and with equal probability. Instead of setting up the moving speed at which users go to the location, we let each user arrive at the contact location immediately at a random time. Each user randomly engages with another user present in the same location for a randomized number of times at a random time. We use minutes as the smallest time unit.

Before the starting of the model each day, each user generates the DTK of the day and calculates the user's real RPI throughout the day. After the model starts, the simulated system randomly determines where a user is going. After the location is determined, this user randomly communicates with any other users in the current location for a randomized number of times. Each communication lasts $t$ minutes. In this process, the two users who communicate with each other will exchange the RPI of the current time slot and perform data processing according to the \texttt{CoAvoid} algorithm. At the end of the day, the number of real contacts of patients will be calculated, and contacts were traced according to the \texttt{CoAvoid} procedures. Each user will be labeled as healthy, suspected, or sick. The required tracing data will be uploaded when a user is determined to be sick. Each user will download the patient’s anonymous data for contact verification. After verification using the \texttt{CoAvoid} rules, if a user confirms contact with a patient, it will be marked as suspected. Based on a pre-set infection rate, suspected cases are marked as sick on a daily basis and will be further utilized for model simulations on the following day.

According to the quarantine time of COVID-19, this paper set up a 14-day simulation experiment, with Xi'an, Shaanxi Province, China as the simulation system’s geographic environment. We utilized the GPS location information of Xi'an to help simulate the places for user contacts. We take the minimum RPI exchange time as the contact time unit. The infection rate is set to be 100\%, which means if a user has any contact with a patient, this user will be infected. The number of actual patients are calculated according to the simulation system, and the number of contacts on the day is calculated according to the \texttt{CoAvoid} algorithm. Android simulators generate the broadcast data used in our experiment according to official documents of GAEN and DP-3T, which simulate broadcast information for the user device. Our edge server is a machine running a Windows 10 operating system with an Intel (R) Core (TM) i3-8100 CPU and a 16GB, which is used to manage uploaded data.

\subsubsection{Privacy Enhancement}

Part of the information broadcasting by a patient during a period is shown in Table \ref{tab:broadcasting}, we can figure that the data contact between users is completely random.
Table \ref{tab:receive} demonstrates the data recorded by the patient's mobile device before uploading the information. As can be seen from the table, since the RPI generated by the contacts and each person is completely random, and the geographical location is encrypted, it is impossible to track a user who has been contacted only through the information collected and broadcast to the device.
Random RPIs and processed GPS information can better protect patients' privacy. 

 \begin{table}[htp]
    	\renewcommand\arraystretch{1.5}    
    	\tabcolsep=2pt
    	\footnotesize
    	\begin{floatrow}
    		\ttabbox{}{%
    			\begin{tabular}{cc}  
    				\Xhline{1.2pt}	
    				Time&RPI\\
    				\Xhline{1.2pt}  
2020-08-09 09:00:00     &     a7776f642d40a688e5fa8232d588bcdd\\\Xhline{0.4pt}  
2020-08-09 09:15:00     &     7a34dbb03355ef6e4cc8382002d86a4c\\\Xhline{0.4pt}  
2020-08-09 09:30:00     &     601d6118b8f2900699c6641bf2eecfa5\\

    				\Xhline{1.0pt}
    			\end{tabular}
    			\caption{Broadcasting logs of patient}
    			\label{tab:broadcasting}}
    	\end{floatrow}
    \end{table}

       \begin{table}[htp]
    	\renewcommand\arraystretch{2.2}    
    	\tabcolsep=2pt
    	\footnotesize
    	\begin{floatrow}
    		\ttabbox{}{
    			\begin{tabular}{ccc}  
    				\Xhline{1.2pt}	
    				Time&RPI&GPS Information\\
    				\Xhline{1.2pt}  
\makecell[c]{2020-08-09\\09:04:00}   &\makecell[c]{f20f8ec68b7ec16\\2427850607b93e5a5} &\makecell[c]{2c05f58240d84224cf0a\\ce831674735314ed655524\\13c8e7f1a9748873d794c8}\\ \Xhline{0.4pt} 
\makecell[c]{2020-08-09\\09:06:00}   &\makecell[c]{f20f8ec68b7ec16\\2427850607b93e5a5} &\makecell[c]{2c05f58240d84224cf0a\\ce831674735314ed655524\\13c8e7f1a9748873d794c8}\\\Xhline{0.4pt} 
\makecell[c]{2020-08-09\\10:40:00}   &\makecell[c]{ae16fac366df90d\\3bca294c9f0a19ba7} &\makecell[c]{2cf33e200fd54585ff8a\\9e24d91e3f93bb72cbb38\\a916792bb1b2bb7fd639cb2}\\\Xhline{0.4pt} 
\makecell[c]{2020-08-09\\10:42:00}   &\makecell[c]{0b13aece18ea9da\\487ea68c54804e1dd} &\makecell[c]{2cf33e200fd54585ff8a\\9e24d91e3f93bb72cbb38\\a916792bb1b2bb7fd639cb2}\\

    				\Xhline{1.0pt}
    			\end{tabular}
    			\caption{Exchange logs of patient}
    			\label{tab:receive}}
    	\end{floatrow}
    \end{table}

The experiment simulated the process of filtering the information uploaded by patients. The experiment demonstrates that the number of RPIs uploaded by users is greatly reduced by filtering the information and interrupting the temporal order of RPIs when they are broadcast. When the attacker conducts a privacy analysis attack, the patient's trajectory can be determined by matching the information collected by the attacker with the patient's RPI on the server. However, the filtered RPI information does not have significant time continuity and removes some information that no other users have been contacted.
Thus, the attacker can only obtain a small number of information about the historical locations of patients, making it more challenging to analyze privacy.

\begin{figure}[htbp]
	\centering
\subfloat[]{\includegraphics[width=1.6in]{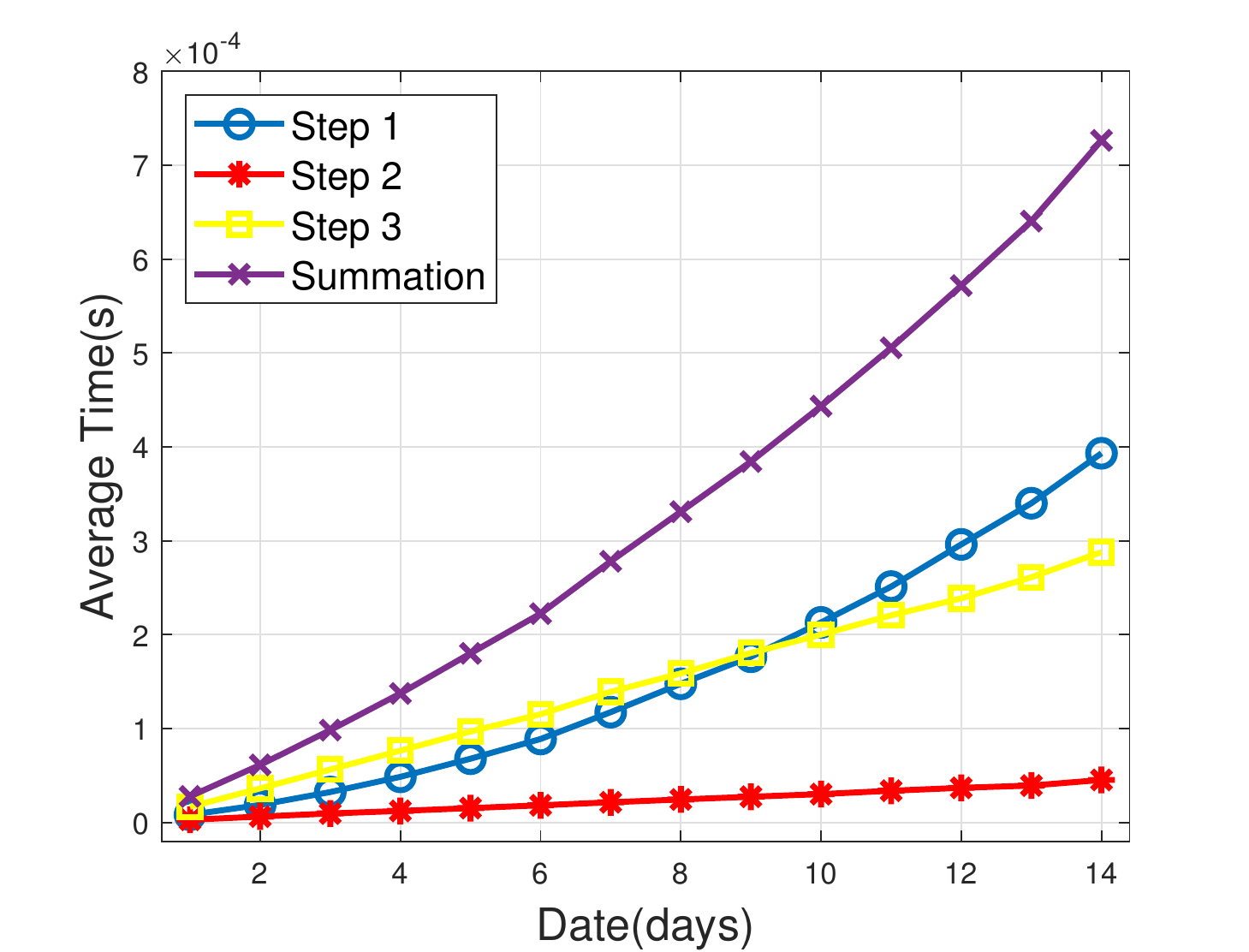}}
	\label{fig:timeconsumptionleft}
	\hfill
	\subfloat[]{\includegraphics[width=1.6in]{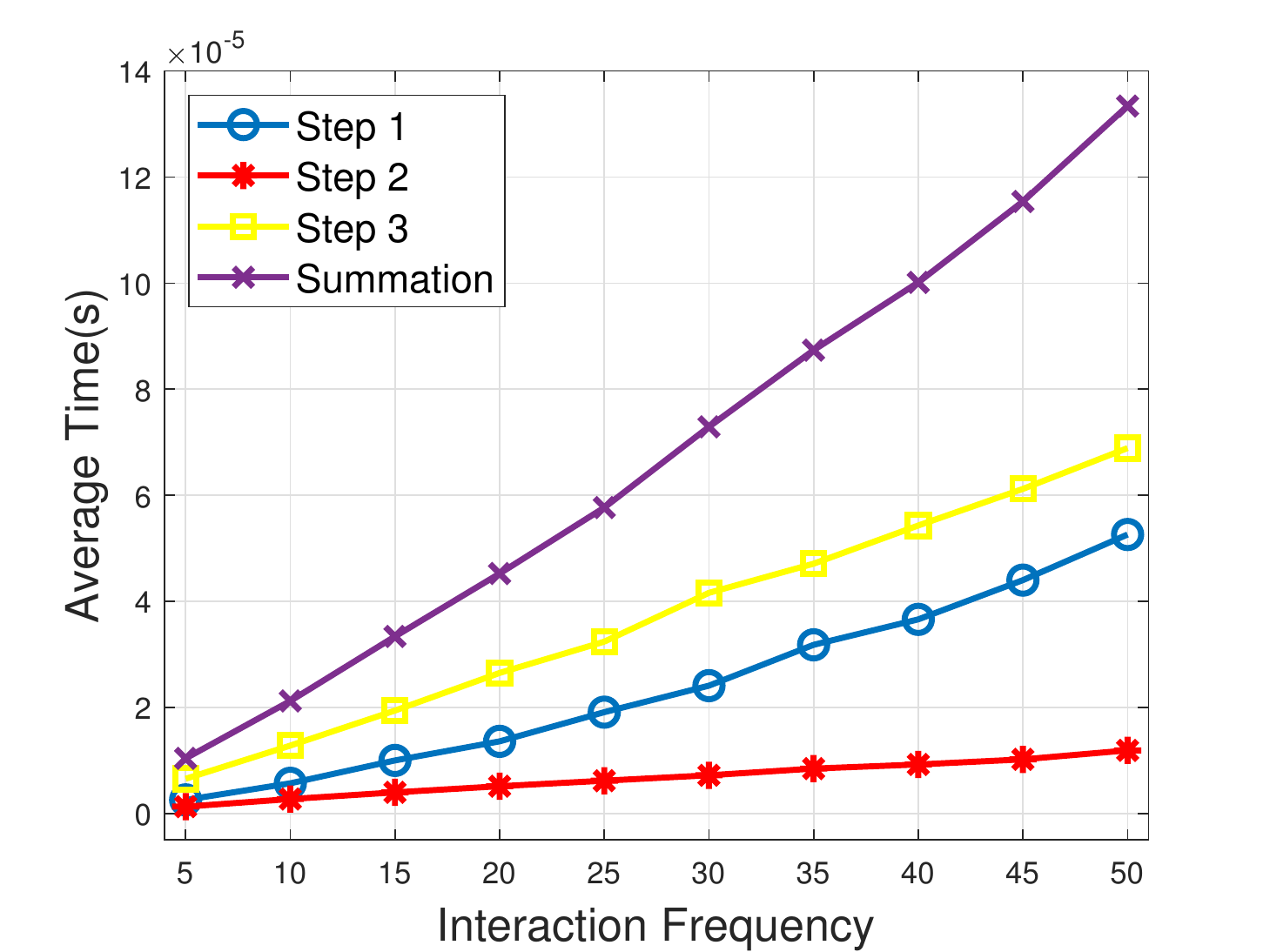}}
	\label{fig:timeconsumptionright}
	\caption{Time consumption for RPI recombination. (a) Date-related time consumption with 100 groups of users contacting at each time unit. (b) Seven-day time consumption with different interaction frequencies.}
	\label{fig:privacy_time_consumption}
\end{figure}

The RPI information uploaded by the patient needs to be processed in three steps. The time consumption of these three steps is demonstrated in Figure \ref{fig:privacy_time_consumption}.
Figure \ref{fig:privacy_time_consumption}(a) illustrates the time consumption for user screening on different numbers of days, where we let 100 groups of users contact each other at each time unit. In Figure \ref{fig:privacy_time_consumption}(b), we set up a seven-day simulation experiment to test the time consumption for user screening at different contact frequencies.
Figure \ref{fig:privacy_time_consumption}(a) and Figure \ref{fig:privacy_time_consumption}(b) show that the time consumption of these three steps increases as date and interaction frequency increases. However, it takes very little time. Users generate more contact data when the proportion of patients remains unchanged. Despite the most data to be calculated, \texttt{CoAvoid} can still finish the two processes within 0.0042s. This feature helps medical staff to obtain patient information in a little time.

After receiving all patient data, the experiment simulated that the server will obfuscate the default storage order to ensure that adjacent RPIs do not belong to the same patient before other users download all the data.

\begin{figure}[htbp]
	\centering
\subfloat[]{\includegraphics[width=1.7in,height=1.4in]{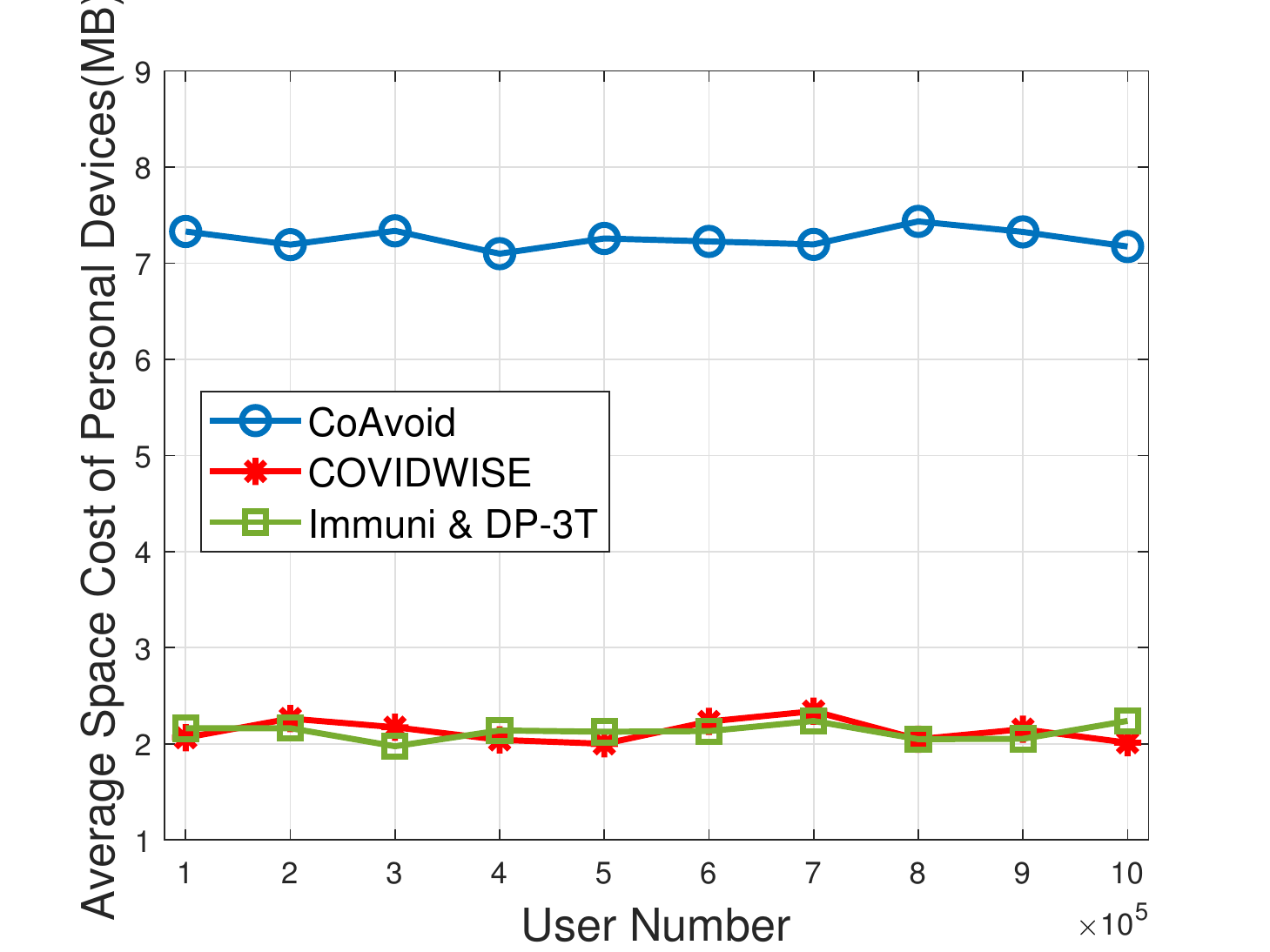}}
	\label{fig:privacy_space_consumption_1}
	\hfill
	\subfloat[]{\includegraphics[width=1.7in,height=1.4in]{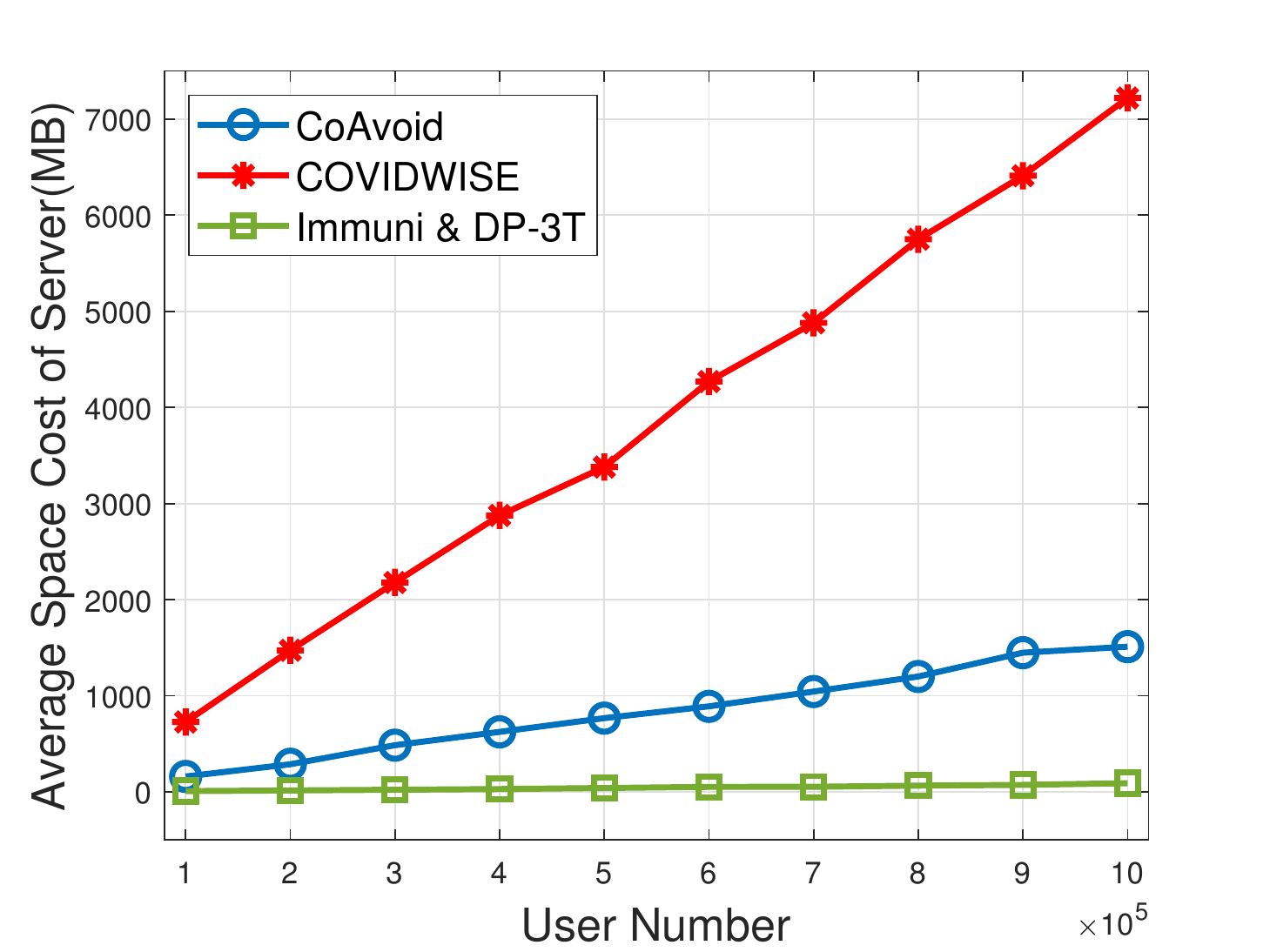}}
	\label{fig:privacy_space_consumption_2}
	\caption{Space consumption comparison. (a) Space consumption of users. (b) Space consumption of server.}
	\label{fig:privacy_space_consumption}
\end{figure}

To evaluate the space consumption for storing interactive information in \texttt{CoAvoid}, we ran different contact tracing approaches on both users' devices and servers with the same number of simulation days and contact frequency (as demonstrated in Figure \ref{fig:privacy_space_consumption}). Figure \ref{fig:privacy_space_consumption}(a) illustrates the space consumption on users’ devices. We can see that the other three distributed tracing applications remain stable . The space consumption of COVIDWISE is 2.1 MB while the space consumption of Immuni and DP-3T is 1.9-2.2 MB. As for \texttt{CoAvoid}, it takes 7.1-7.3 MB for data storage. Figure \ref{fig:privacy_space_consumption}(b) illustrates the space consumption on the server. \texttt{CoAvoid} saves 77.5\% - 79\% space compared with COVIDWISE. As the amount of information uploaded by each patient is fixed in COVIDWISE, its server space consumption is proportional to the number of patients. Besides, in \texttt{CoAvoid}, the amount of information uploaded by each patient is determined by their activities and will be filtered before uploading. Therefore, \texttt{CoAvoid}’s server space consumption is not strictly linear increasing, and the growth is much slower than COVIDWISE. Immuni and DP-3T upload very little information to servers because the DTK they upload can greatly reduce the space.

Compared to other methods, our approach requires more space, especially on users’ devices, to defend against wormhole attacks in any scenario, while other approaches are vulnerable to wormhole attacks. Still, for personal devices, the space consumption of our approach is within an acceptable range, which is only slightly higher than other solutions. Although other solutions take up relatively less storage space, the information uploaded by other solutions exposes more sensitive user information. Our approach offers more comprehensive privacy protection and can conduct tracing tasks more securely.

\begin{figure}[htbp]
	\centering
\subfloat[]{\includegraphics[width=1.7in,height=1.4in]{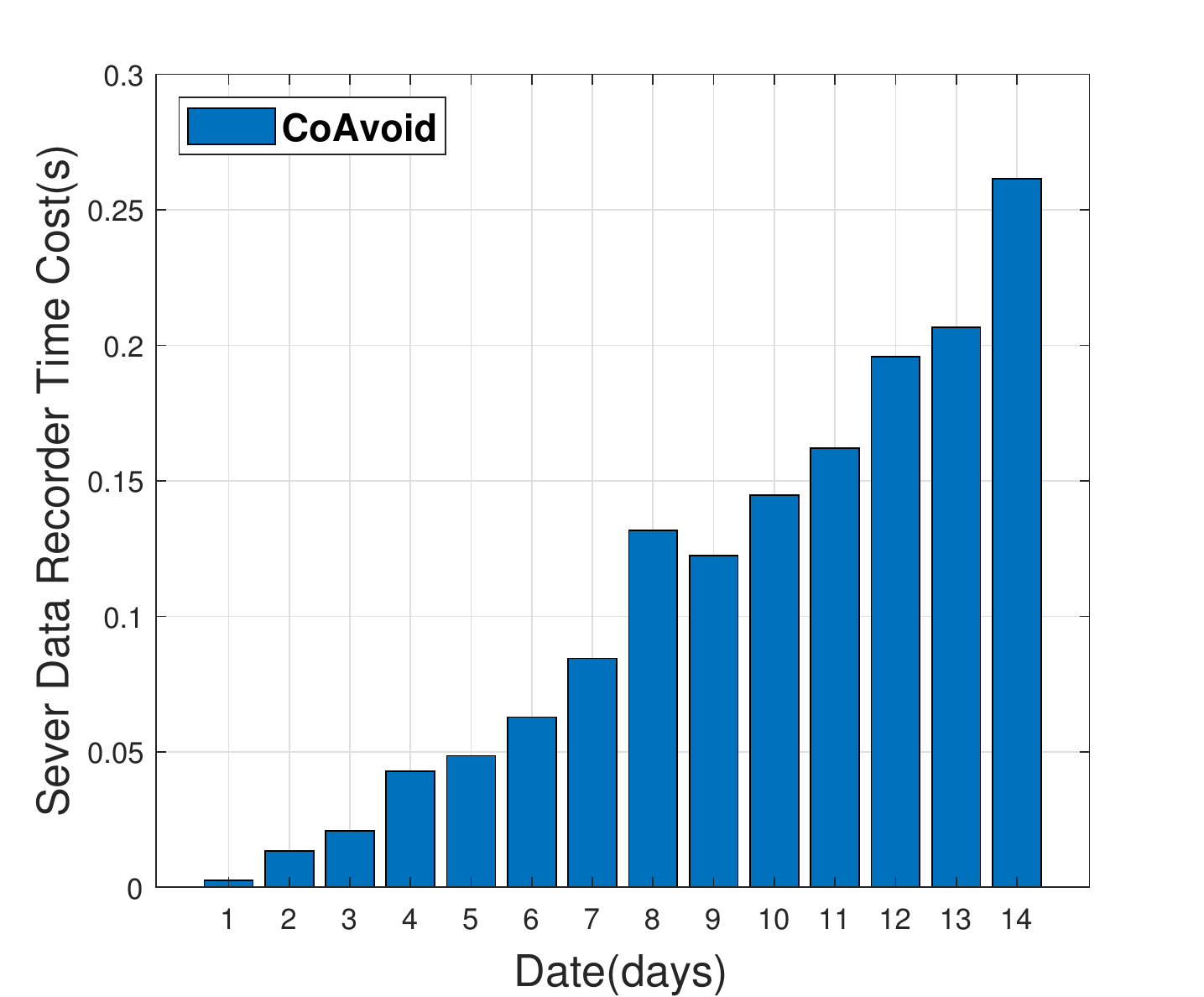}}
	\label{fig:privacy_reorder1}
	\hfill
	\subfloat[]{\includegraphics[width=1.7in,height=1.4in]{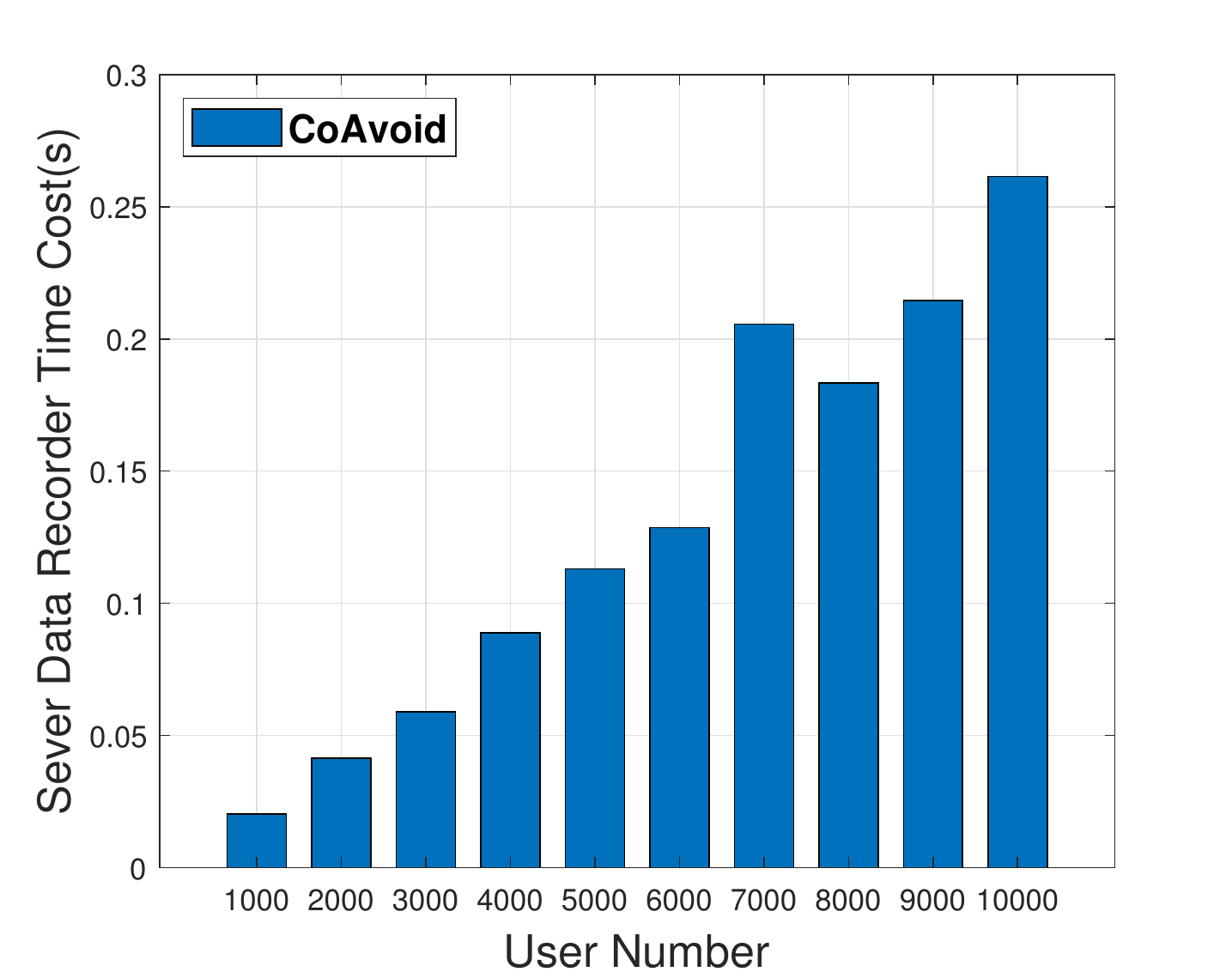}}
	\label{fig:privacy_reorder2}
	\caption{Server data reorder time cost. (a) Reorder time cost with day time for 10,000 people. (b) 14-day reorder time cost with user number.}
	\label{fig:privacy_reorder}
\end{figure}

Figure \ref{fig:privacy_reorder} demonstrates the time costs for the server when dealing with reordering the data. Figure \ref{fig:privacy_reorder}(a) tests the server obfuscated data time consumption on different days for 10,000 people, and Figure \ref{fig:privacy_reorder}(b) sets up the server obfuscated data time consumption in 14 days for a different number of people. The time cost can increase with the number of users or days. However, the burden on the server is not significant. The result demonstrates the time consumption is only 0.261s, even on the 14th day or with 10000 users.

\subsubsection{Attack Prevention}

To evaluate the wormhole attack resistance of \texttt{CoAvoid}, we generated the false BLE messages propagated by the attacker and test whether these messages can be identified during the message verification step on users’ devices. Theoretically, such false messages will be identified by verifying the location information recorded by the user device that receives the patient's BLE message.

\begin{figure}[htp]
	\centering
	\includegraphics[width=3.2in]{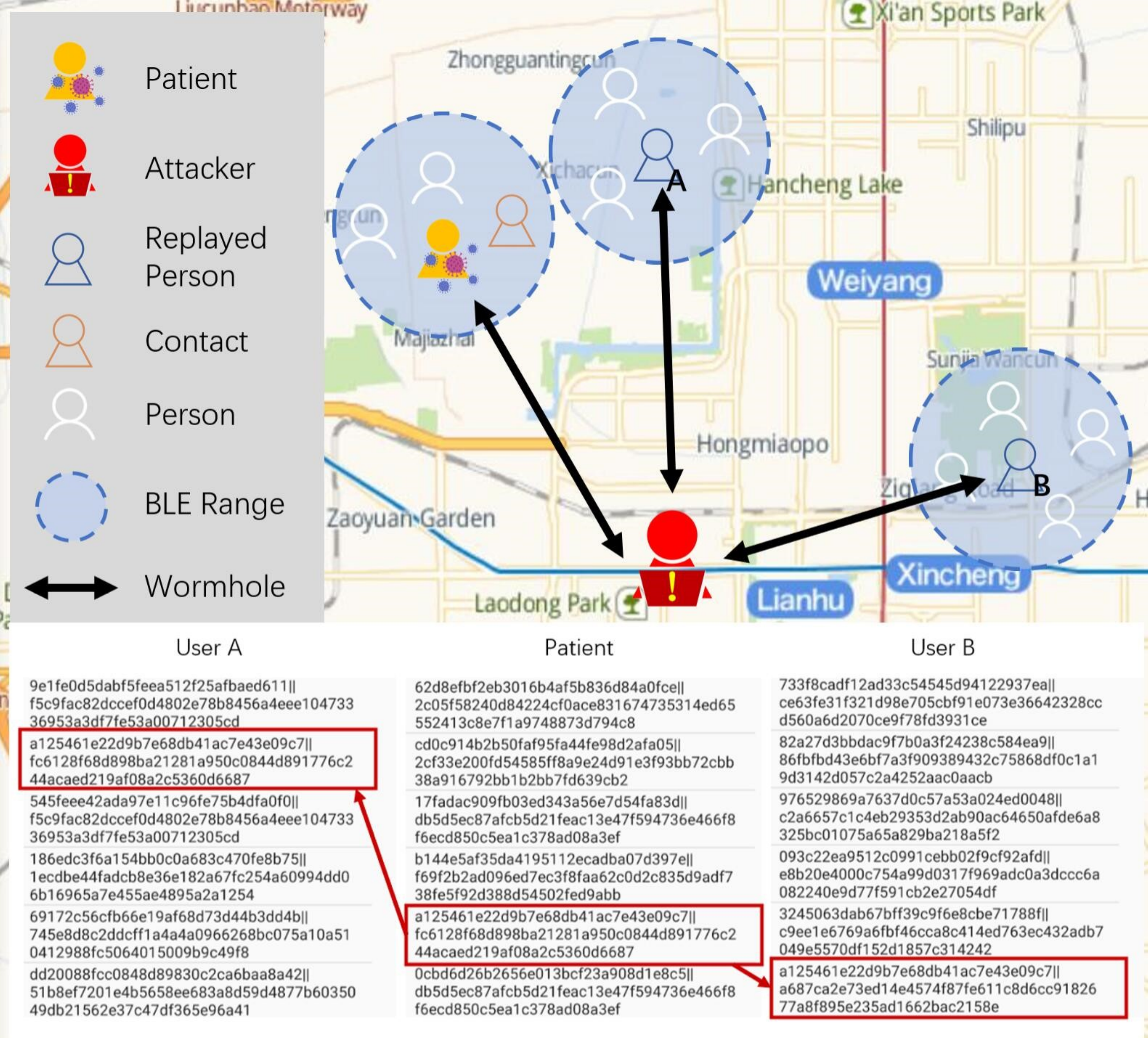}
	\caption{Operation model of the wormhole attack in the city of Xi’an and the corresponding log information.}
	\captionsetup{font={scriptsize}}
	\label{fig:wattack}
\end{figure}

We built a multi-location wormhole for forwarding and sending fake BLE messages. Our experimental links the physical locations of Xi'an, Xianyang, and some other cities. The attack within Xi'an city is demonstrated in Figure \ref{fig:wattack}. The attacker receives and propagates Bluetooth messages sent by users around the city and transmit the messages to other attackers in other locations for broadcast. Each malicious device is a device that can send and receive Bluetooth messages, so the attackers’ communications are operating in a multi-way fashion.

Our experiments show that our approach can successfully resist wormhole attacks. For example, the same BLE message appears in different devices in different cities, indicating a successful wormhole attack. However, the user device does not determine this as a patient contact because the device will verify the user's location information while contact tracing, which eliminates the effects of false BLE information on the devices’ judgments.

Figure \ref{fig:wattack} also illustrates the log information of users and patients in two different locations that were attacked by the wormhole, where user A is a replayed user in the coarse precision range and user B is a relayed user outside the coarse precision range. The data in the log file is presented in the format $RPI||f_{SHA256}(f_{H3}(l))$. It can be seen that because of the presence of the patient's Bluetooth information in the device logs, both users are under wormhole attack. Listing \ref{lst:listing} demonstrates the log extracts from the two user devices mentioned above. After both user devices identify the patient's BLE information, coarse precision location verification is required for both user devices. The log result will show "Wormhole Attack" in line 1 because the user B device matches the same patient information, but the location information does not match. In line 2, if the location information of user A matches, log result will show "Correct", and then the fine-grained screening will be performed. In the fine-grained verification, the patient and the user will push encryption parameters to each other and check whether the data calculated by the user meets the judgment condition. In line 3, if the user finally gets a positive value, it will be judged as a normal user under a wormhole attack and the log result will show "Wormhole Attack".

\begin{lstlisting}[   
	language= ,
	label={lst:listing},
	caption={An example of location verification.},
	backgroundcolor = \color{yellow!10}, 
	basicstyle=\ttfamily,
	breaklines=true,
	keywordstyle=\color{NavyBlue},
	morekeywords={INFO,},
	emph={Wormhole,Attack},
    emphstyle=\color{Rhodamine},
    commentstyle=\itshape\color{black!50!white},
    emph={[2]Final,Correct,U,P}, emphstyle={[2]\color{PineGreen}},
    stringstyle=\bfseries\color{PineGreen!90!black},
    columns=flexible,
    numbers=left,
    linewidth=1.03\linewidth,
    xleftmargin=1em,
    numbersep=0.7em,
    numberstyle=\footnotesize
]  
Location Verification[1]: [INFO] [P] fc6128f68d898ba21281a950c0844d891776c2   44acaed219af08a2c5360d6687 [U] a687ca2e73ed14e4574f87fe611c8d6cc91826 77a8f895e235ad1662bac2158e [Wormhole Attack]
Location Verification[1]: [INFO] [P] fc6128f68d898ba21281a950c0844d891776c2 44acaed219af08a2c5360d6687 [U] fc6128f68d898ba21281a950c0844d891776c2 44acaed219af08a2c5360d6687 [Correct]
Location Verification[2]: [INFO] [Final] 3.8422948129866016e+197 [Wormhole Attack]
\end{lstlisting}

\subsubsection{Server Data Storage Comparison}

\begin{figure*}[htbp]
	\centering
	\subfloat[]{ %
		\label{fig:performance_8_1} 
		\includegraphics[width=1.7in]{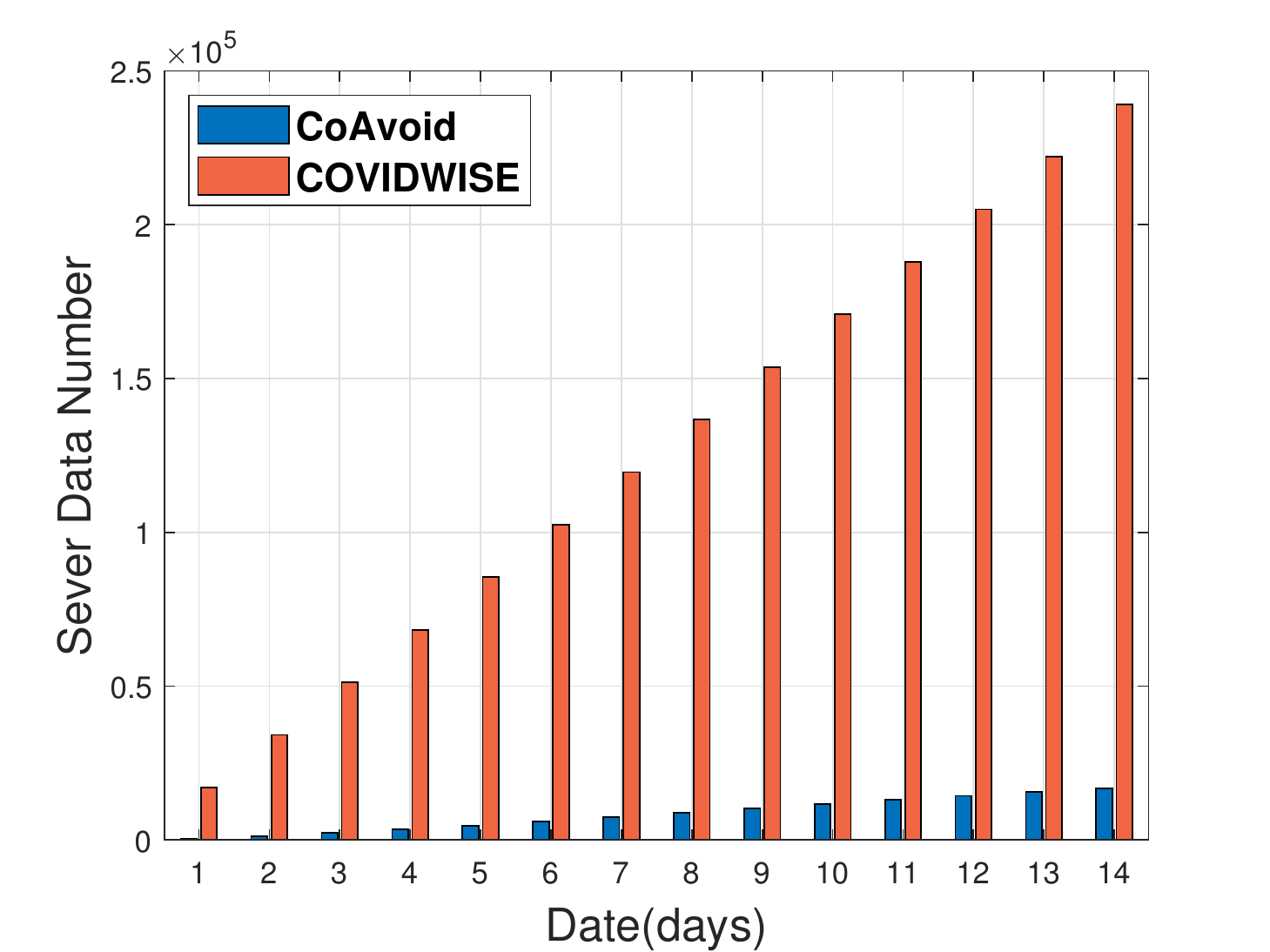}}
	\subfloat[]{
		\label{fig:performance_8_2} %%
		\includegraphics[width=1.7in]{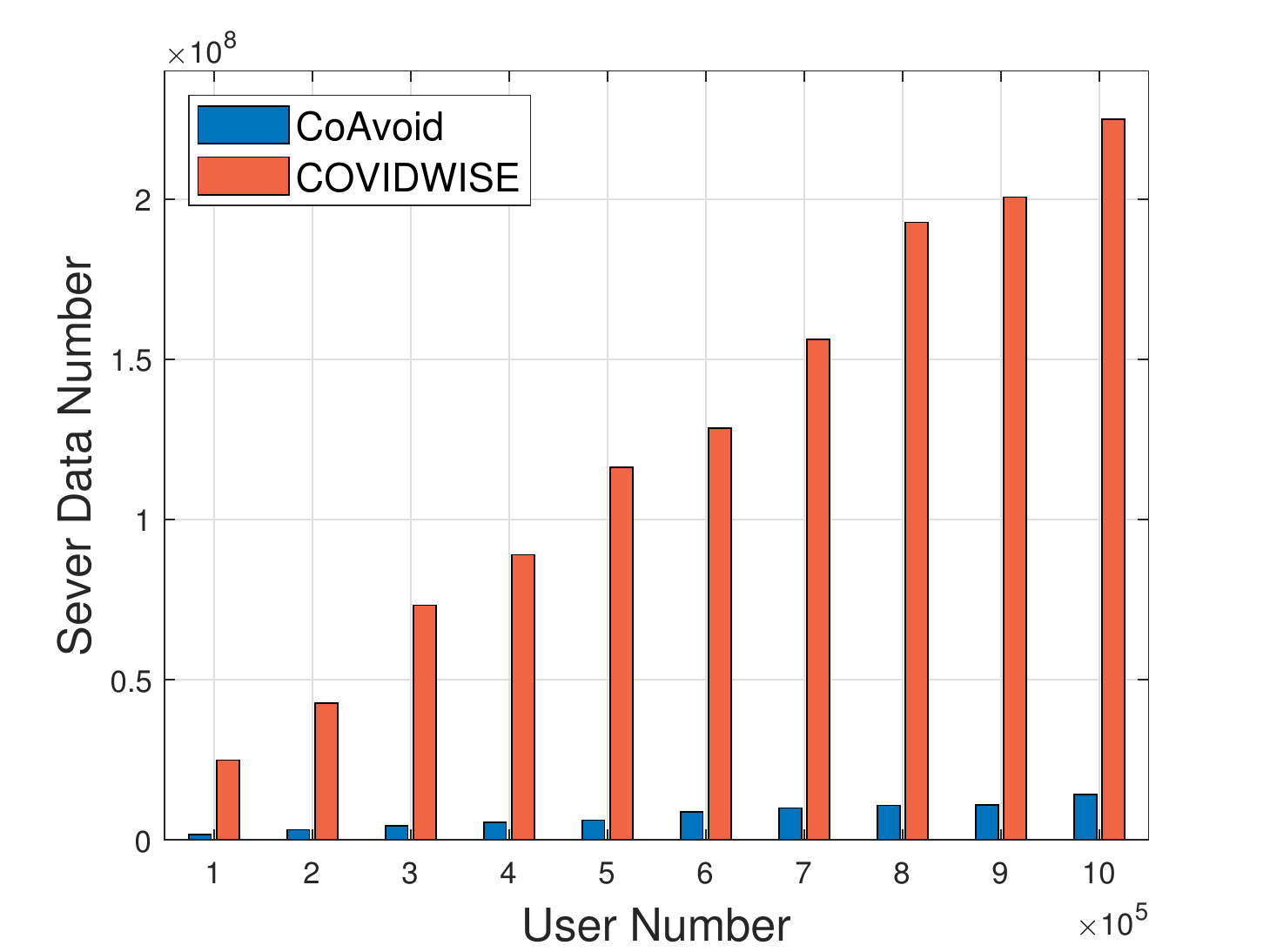}}
	\subfloat[]{
		\label{fig:performance_8_3} %%
		\includegraphics[width=1.7in]{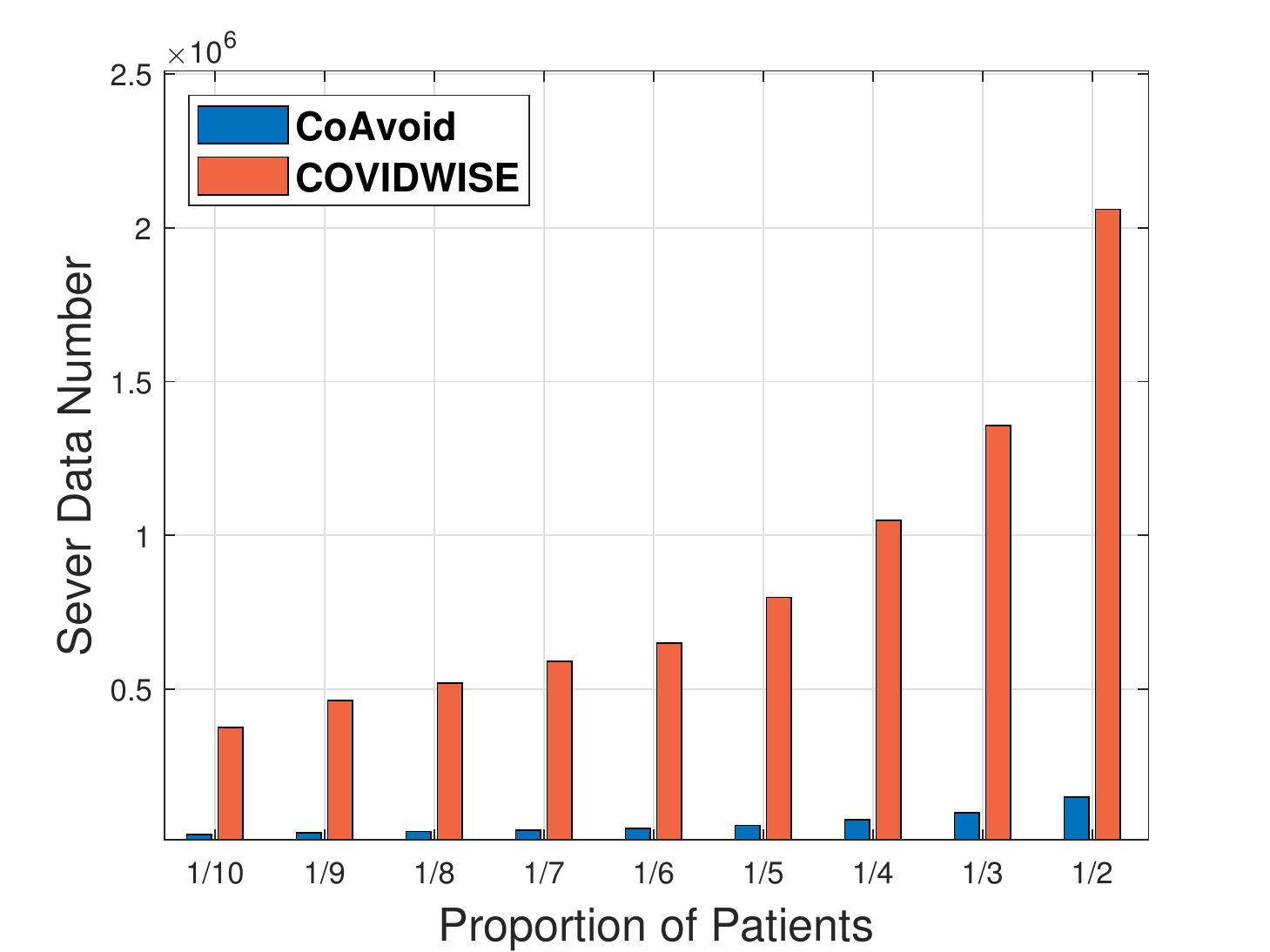}}
	\subfloat[]{
		\label{fig:performance_8_4} %%
		\includegraphics[width=1.7in]{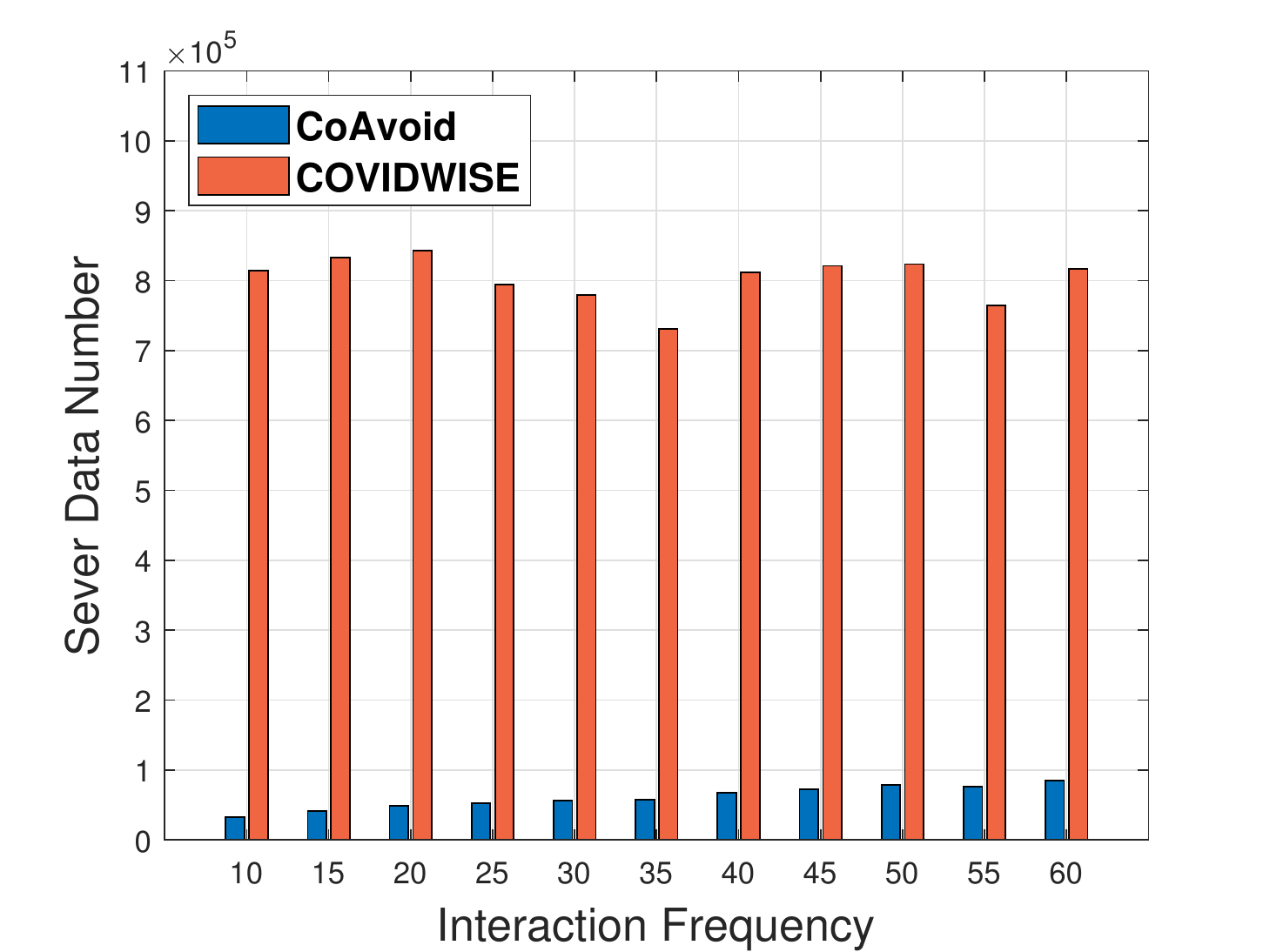}}
	
	\caption{Server data storage comparison between COVIDWISE and CoAvoid. (a) Server data with days for 10,000 people. (b) 14-day server data with different numbers of users. (c) Server data with patients for 10,000 people in 14 days. (d) Server data with interaction frequecny of 10,000 people}
\label{fig:4store} %%
\end{figure*}

When a patient uploads their interaction data to the edge server, the user device needs to regularly download the patient’s data to verify current traces with it. If a coarse-granular record of interactions with the patients exists in the user device, then the user may have contact with the disease. The data size uploaded to the server is an important factor that affects the system's efficiency. We thus evaluated and compared the data volume uploaded by patients for coarse-grained comparisons in different scenarios, as demonstrated in Figure \ref{fig:4store}. Compared to COVIDWISE, which uploads all the contact data of patients in the past 14 days to the server, \texttt{CoAvoid} only uploads contact data generated when contacting other users.
Figure \ref{fig:4store}(a) demonstrates the comparison of server data volume for 10,000 people on different days. The data in the \texttt{CoAvoid} server grows as the date increases, but its data is only 3\% - 8\% of COVIDWISE, as demonstrated in Figure \ref{fig:4store}(a).

In addition, we selected 100,000 to 1,000,000 users to evaluate the amount of data uploaded to the server from the perspective of the population. We simulated different cases with the same interaction rule for a period of 14 days.
Figure \ref{fig:4store}(b) illustrates the server data volume. As the user number increases, the population density and contact frequency will also increase, which makes healthy people more likely to be infected. Compared with COVIDWISE, \texttt{CoAvoid} can reduce 92\% - 94\% of uploading data.

Figure \ref{fig:4store}(c) illustrates the comparison of server data volume for 10,000 people in 14 days with different patient proportions and interaction frequencies. We can see that the server data volume of COVIDWISE and \texttt{CoAvoid} increases while the number of positive patients increases.
\texttt{CoAvoid} saves 92.7\% - 92.9\% the amount of data on the server compared with COVIDWISE, which dramatically saves the storage space of the server. The results illustrate that our method can significantly reduce the data on the server, which is beneficial for areas with large population and can effectively utilize the server's storage resources.

In figure \ref{fig:4store}(d), we set up a 14-day experiment with 10,000 users to test the influence of interaction frequency on server data volume of two methods. The amount of server data in our approach increases with the frequency of interactions, while the amount of COVIDWISE data remains essentially constant on the server. Because the amount of data uploaded by each patient of COVIDWISE is inevitable, the number of patients determines the amount of data on the server. In \texttt{CoAvoid}, the data uploaded by patients is filtered out according to the contact information. The growth of contact data and the increase of screened information of patients uploaded to the server led to the growth of server data. In the experiment, \texttt{CoAvoid}'s server accounted for 10.5\% of COVIDWISE data.

\subsubsection{Verification Time Comparison and Accuracy}

\begin{figure}[htbp]
	\centering
\subfloat[]{\includegraphics[width=1.58in,height=1.4in]{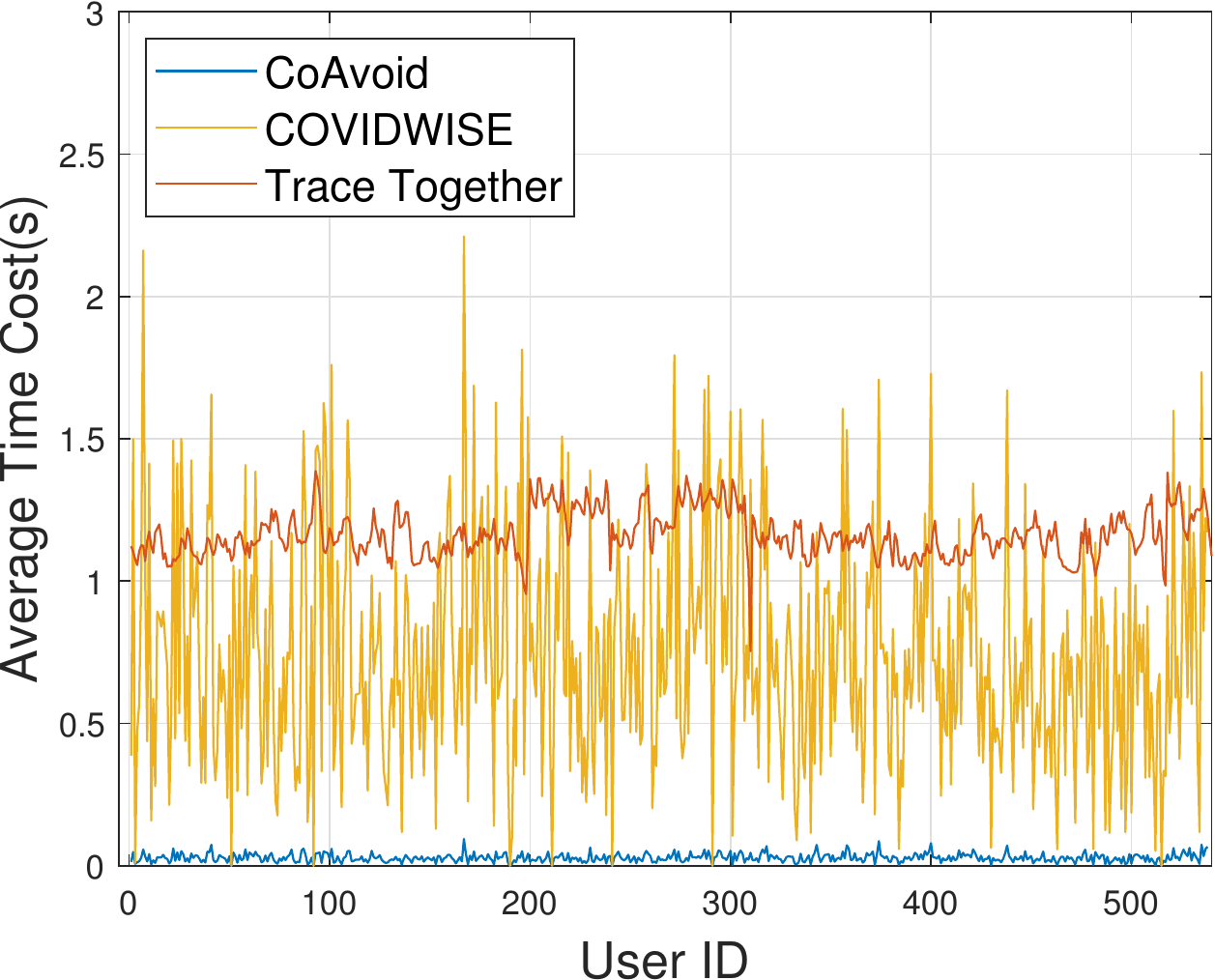}}
	\label{fig:performance_8_5}
	\hfill
	\subfloat[]{\includegraphics[width=1.83in,height=1.5in]{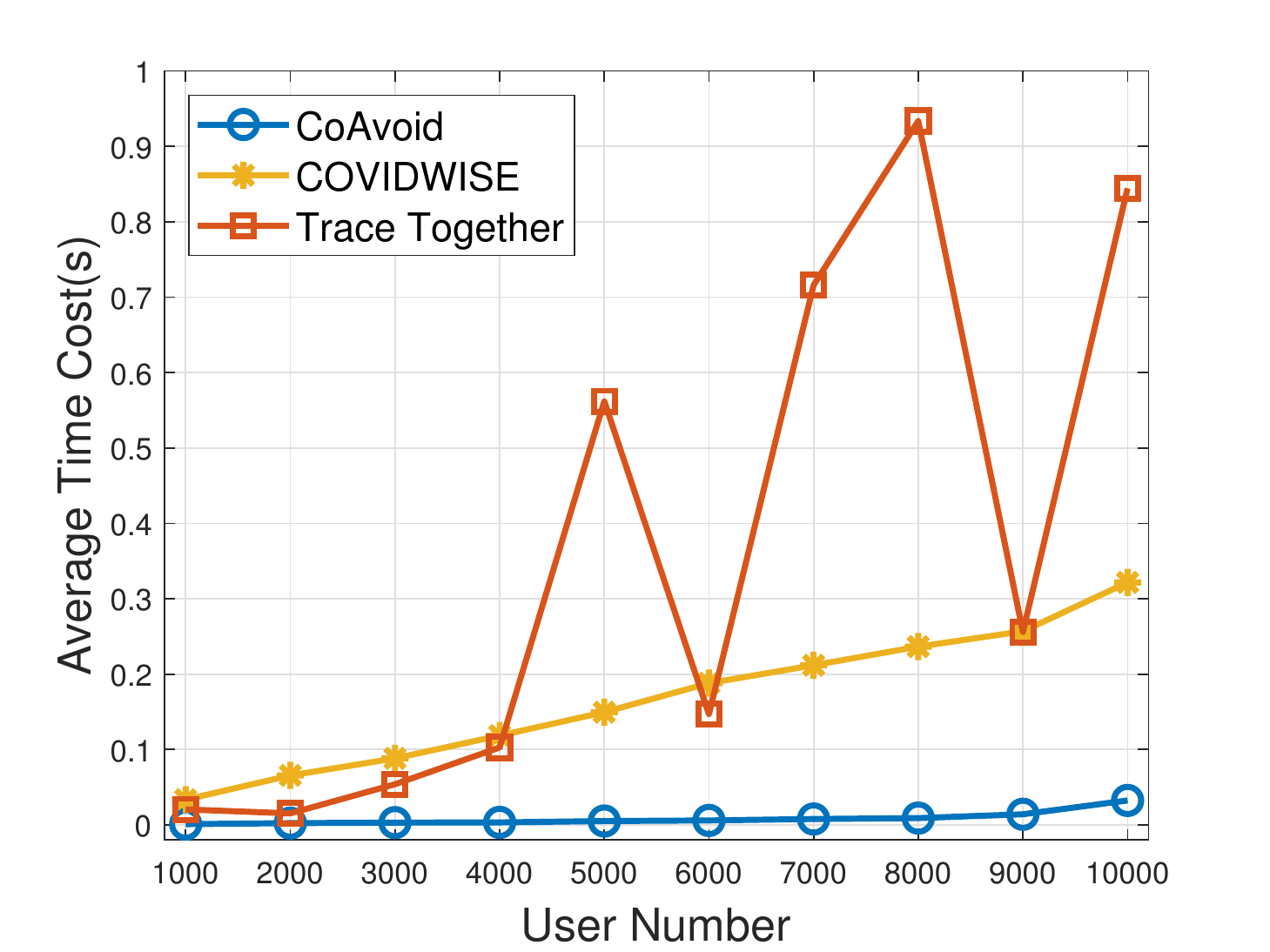}}
	\label{fig:performance_8_6}
	\caption{(a)Verification time comparison with a single user: CoAvoid vs. COVIDWISE vs. TraceTogether. (b)Verification time comparison with different numbers of users: CoAvoid vs. COVIDWISE vs. TraceTogether.}
	\label{fig:4verify}
\end{figure}

Since our scheme uses a lightweight and efficient contact matching algorithm with low computational complexity and communication overhead, it significantly reduces the computation time and burden of the user's device during the verification process, which provides a decent user experience. As demonstrated in Figure \ref{fig:4verify}, we utilize simulated server data to compare the verification time of \texttt{CoAvoid}, COVIDWISE, and TraceTogether. Here, the verification time includes comparing the RPI and GPS information and a fine-grained matching.

Figure \ref{fig:4verify}(a) compares three scenarios about verification time consumption, which is the simulation generated by 540 users. With the same device, TraceTogether takes 1-1.4 seconds for each user in the centralized server to know whether he has contact with patients. In comparison, COVIDWISE takes around 0.15 to 2.21 seconds, and \texttt{CoAvoid} takes less than 0.1 seconds. Our approach incurs a shorter verification time, which reduces the computational burden of user equipment and improves verification efficiency.

Figure \ref{fig:4verify}(b) demonstrates the verification time consumption of different schemes under different numbers of users. For regions with more users, COVIDWISE will generate a large amount of server data. All users' average verification time consumption will be significantly increased, which reduces the verification efficiency. However, the amount of data that is uploaded by \texttt{CoAvoid} in areas with a large number of users is also minimal.
Figure \ref{fig:4verify}(b) demonstrates that in an area of 10,000 users, the average validation time for centralized servers was 0.84 seconds. In contrast, the average validation time was 0.32 seconds for COVIDWISE and 0.028 seconds for \texttt{CoAvoid}. \texttt{CoAvoid} only took 8.7\% of COVIDWISE’s validation time and 3.8\% of the centralized server’s validation time. Therefore, \texttt{CoAvoid} can effectively reduce user authentication time, improving authentication efficiency.

\begin{figure}[htbp]
	\centering
\subfloat[]{\includegraphics[width=1.7in]{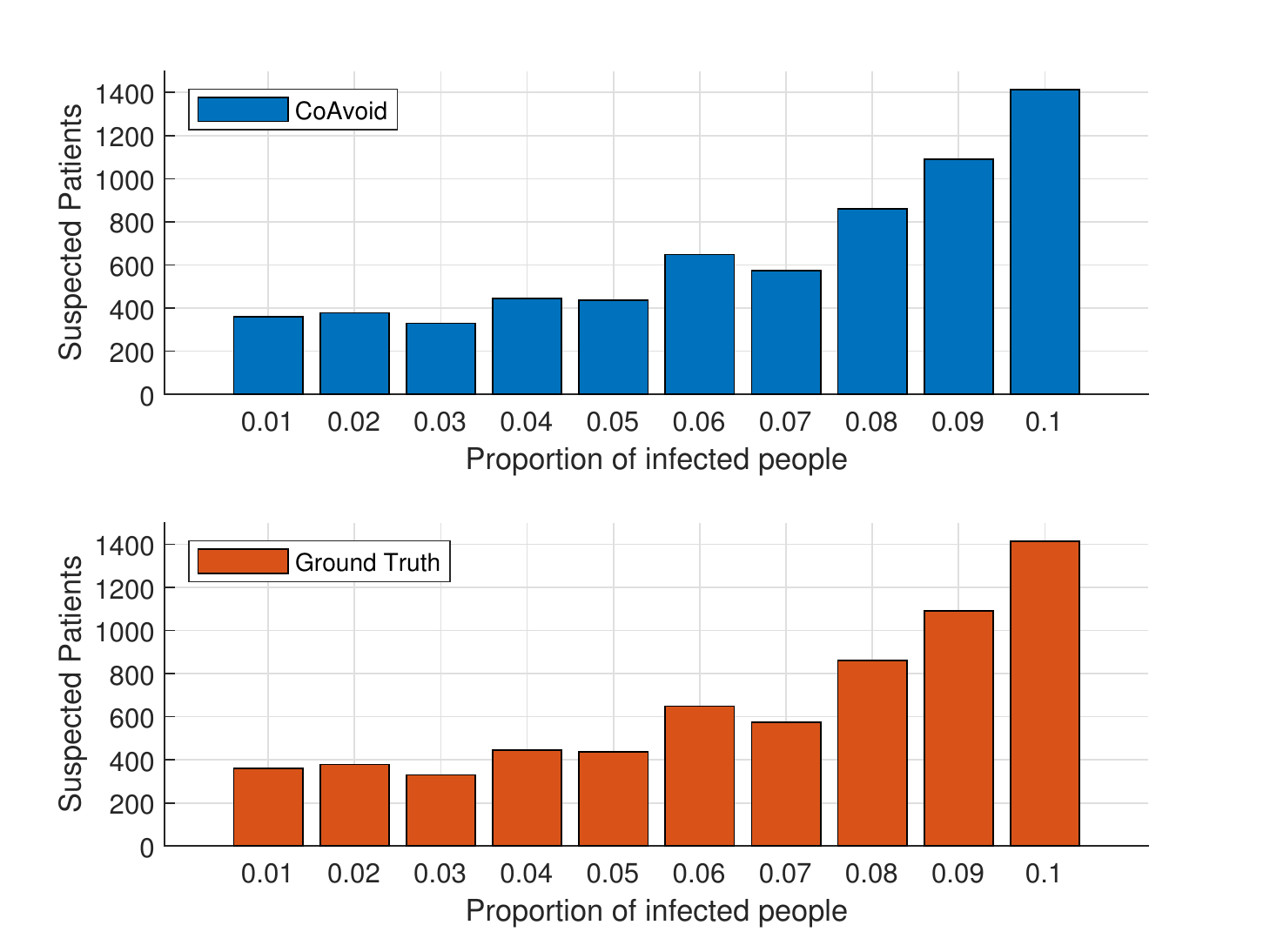}}
	\label{fig:accuracy_verification_of_CoAvoid_left}
	\hfill
	\subfloat[]{\includegraphics[width=1.7in]{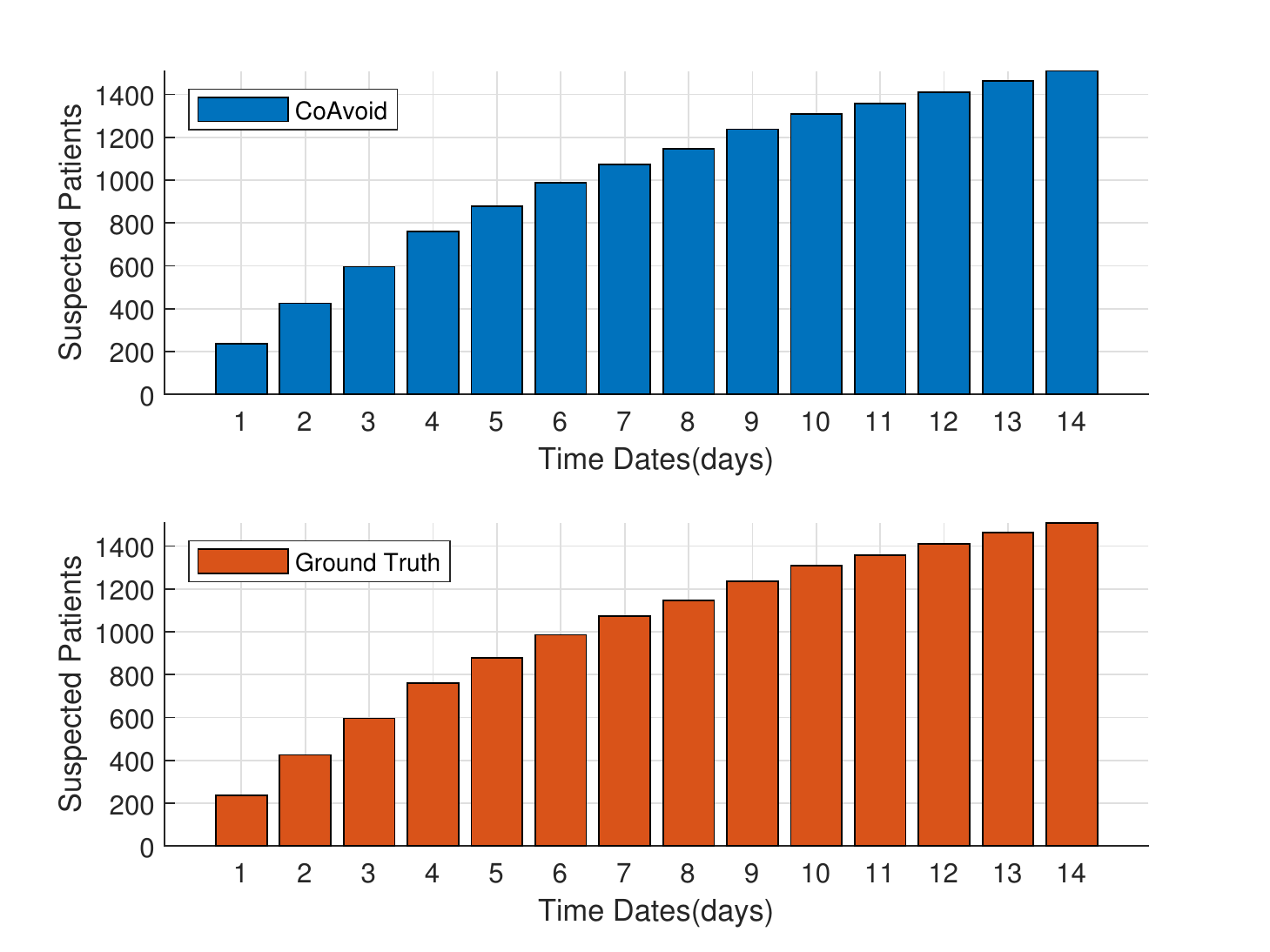}}
	\label{fig:accuracy_verification_of_CoAvoid_right}
	\caption{Accuracy verification of CoAvoid. (a) Accuracy verification with different proportions of infected people within 12 days. (b) Accuracy verification with different numbers of days (the proportion of infected people: 0.1}
	\label{fig:accuracy_verification_of_CoAvoid}
\end{figure}

Figure \ref{fig:accuracy_verification_of_CoAvoid} demonstrates the correlation between the number of contacts detected by \texttt{CoAvoid} and the number of contacts generated by the simulation. The infection rate is set to 100\%, which means if a user has communicated with a patient, this user will be infected. The daily number of actual patients and actual contacts was calculated according to the simulation system, and the number of contacts on the day was calculated according to the \texttt{CoAvoid} algorithm.
Figure \ref{fig:accuracy_verification_of_CoAvoid}(a) demonstrates that \texttt{CoAvoid} had the ability to detect all contacts simulated experimentally in different percentages of patients at 12 days. As demonstrated in Figure \ref{fig:accuracy_verification_of_CoAvoid}(b), \texttt{CoAvoid} can detect all the contact points simulated in the experiment every day as the daily interaction continues when the infected proportion is 0.1. In both experiments, 100\% accuracy is expected, as they validate \texttt{CoAvoid}'s ability to track contacts with complete tracing data as inputs.

\section{Conclusion}
\label{sec:concl}

Contact tracing is the process of identifying those who may have been exposed to someone with a virus, whether COVID-19 or another illness. As the epidemic continues to worsen, the role of contact tracing becomes even more important. To solve the problems in current contact tracing systems, protect user privacy, and resist wormhole attacks, we propose \texttt{CoAvoid}. It is an edge-based contact tracing system based on GAEN API and can achieve a high accuracy on dynamic user trajectory matching with low system and time overhead.
    
By harnessing multiple security designs, such as location verification using GPS and fine-grained matching algorithms, \texttt{CoAvoid} is able to resist both replay and wormhole attacks. To enhance the privacy protection for all users, \texttt{CoAvoid} hides geographic location of users through hashing and fuzzification, filters out uncorrelated contact data before uploading, and obfuscates information uploaded by confirmed patients before storage and analysis. As a result, \texttt{CoAvoid} not only preserves the privacy of low-risk populations, but also makes the public agnostic to high-risk populations' identities and social relationships. Furthermore, as the data to be transmitted and analyzed in \texttt{CoAvoid} system is significantly reduced, our approach can achieve good performance with limited bandwidth and device capacities, enabling it to operate in regions with different levels of development. In addition, benefits from the use of GAEN API, \texttt{CoAvoid} has board hardware and software compatibility.
\section*{Acknowledgments}
This research is funded by National Natural Science Foundation of China (No. 61902291), the founding of Shaanxi Key Laboratory for Network Computing and Security Technology (NCST2021YB-03), the Fundamental Research Funds for the Central Universities (XJS211516), Shaanxi University Science and Technology Association Youth Talent Promotion Project (20210120).

\bibliographystyle{IEEEtran}
\bibliography{reference}
\begin{IEEEbiography}[{\includegraphics[width=1in,height=1.25in,clip,keepaspectratio]{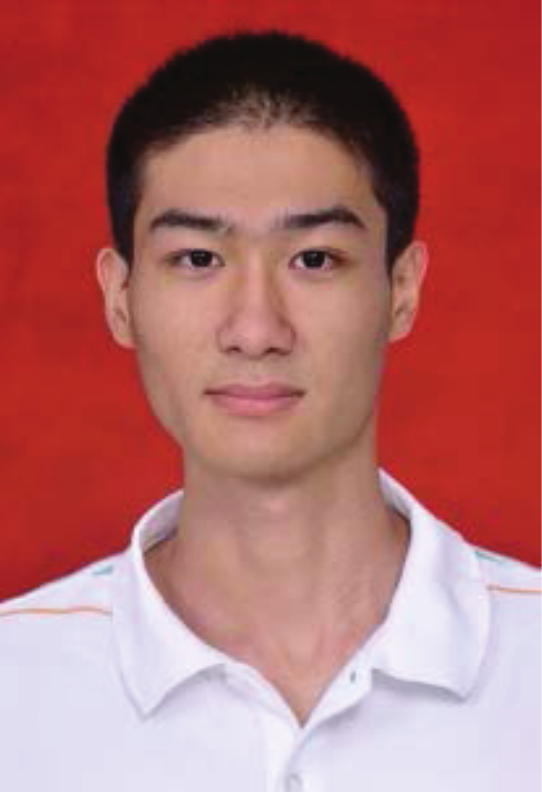}}]{Teng~Li}
 received the B.S. degree in school of computer science and technology from Xidian University, China in 2013, and Ph. D. degree in school of computer science and technology from Xidian University, China in 2018. He is currently an Associate Professor at the school of cyber engineering, Xidian University, China. His current research interests include wireless and mobile networks, distributed systems and intelligent terminals with focus on security and privacy issues.
\end{IEEEbiography}

\begin{IEEEbiography}[{\includegraphics[width=1in,height=1.25in,clip,keepaspectratio]{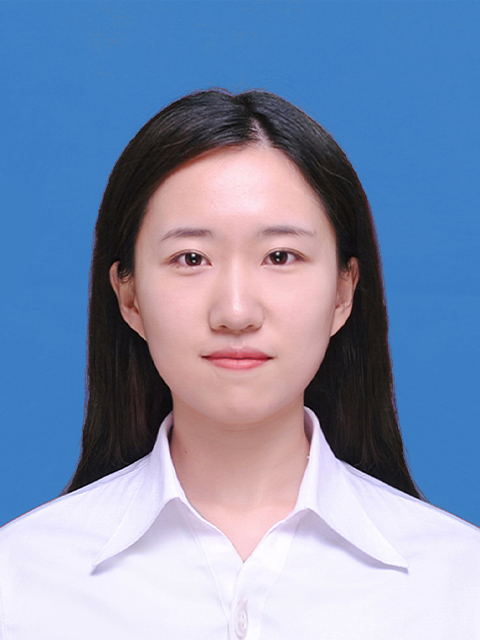}}]{Siwei Yin}
 received the B.S. degree in network Engineering from Xidian University, Shaanxi, China, in 2021. She is currently pursuing the M.S. degree in cyberspace security at Xidian University, Shaanxi, China.
\end{IEEEbiography}

\begin{IEEEbiography}[{\includegraphics[width=1in,height=1.25in,clip,keepaspectratio]{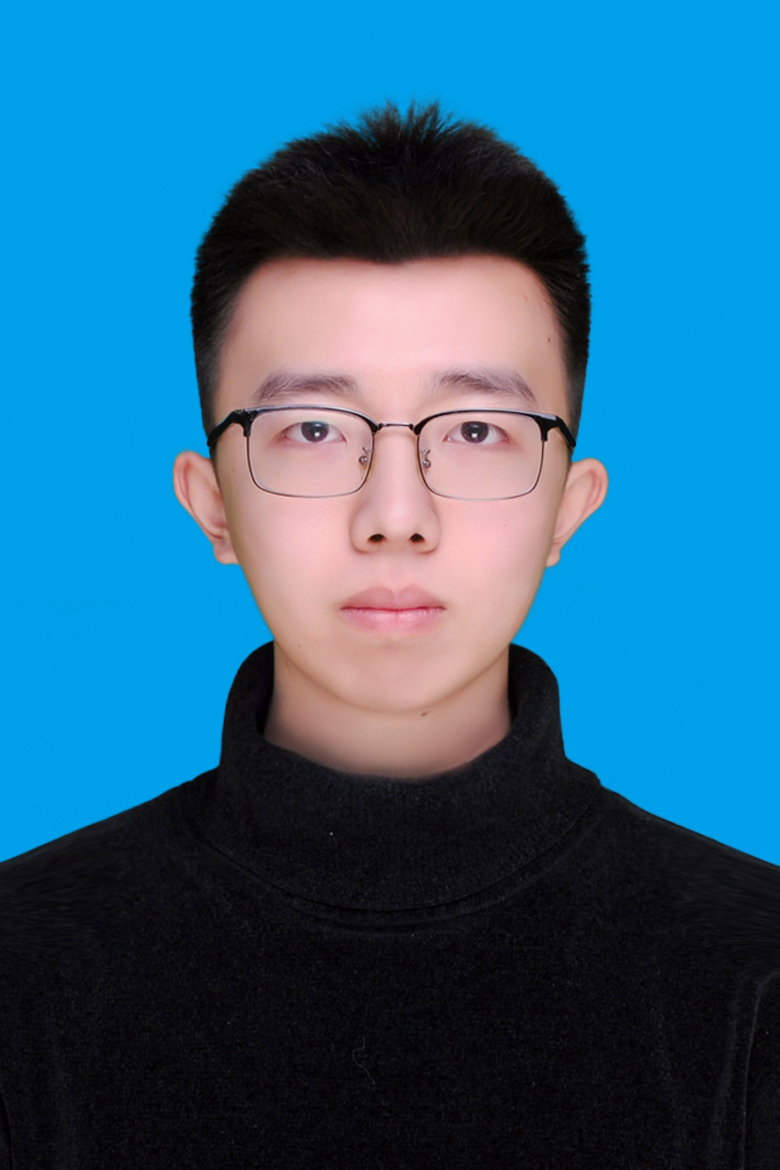}}]{Runze Yu}
 received the B.E. degree from the School of Cyber Engineering, Xidian University, Xi’an, China, in 2021, where he is currently pursuing the master’s degree with the School of Cyber Engineering. His main research interests include network attacks detection, and UAV security and fault detection. 
\end{IEEEbiography}

\begin{IEEEbiography}[{\includegraphics[width=1in,height=1.25in,clip,keepaspectratio]{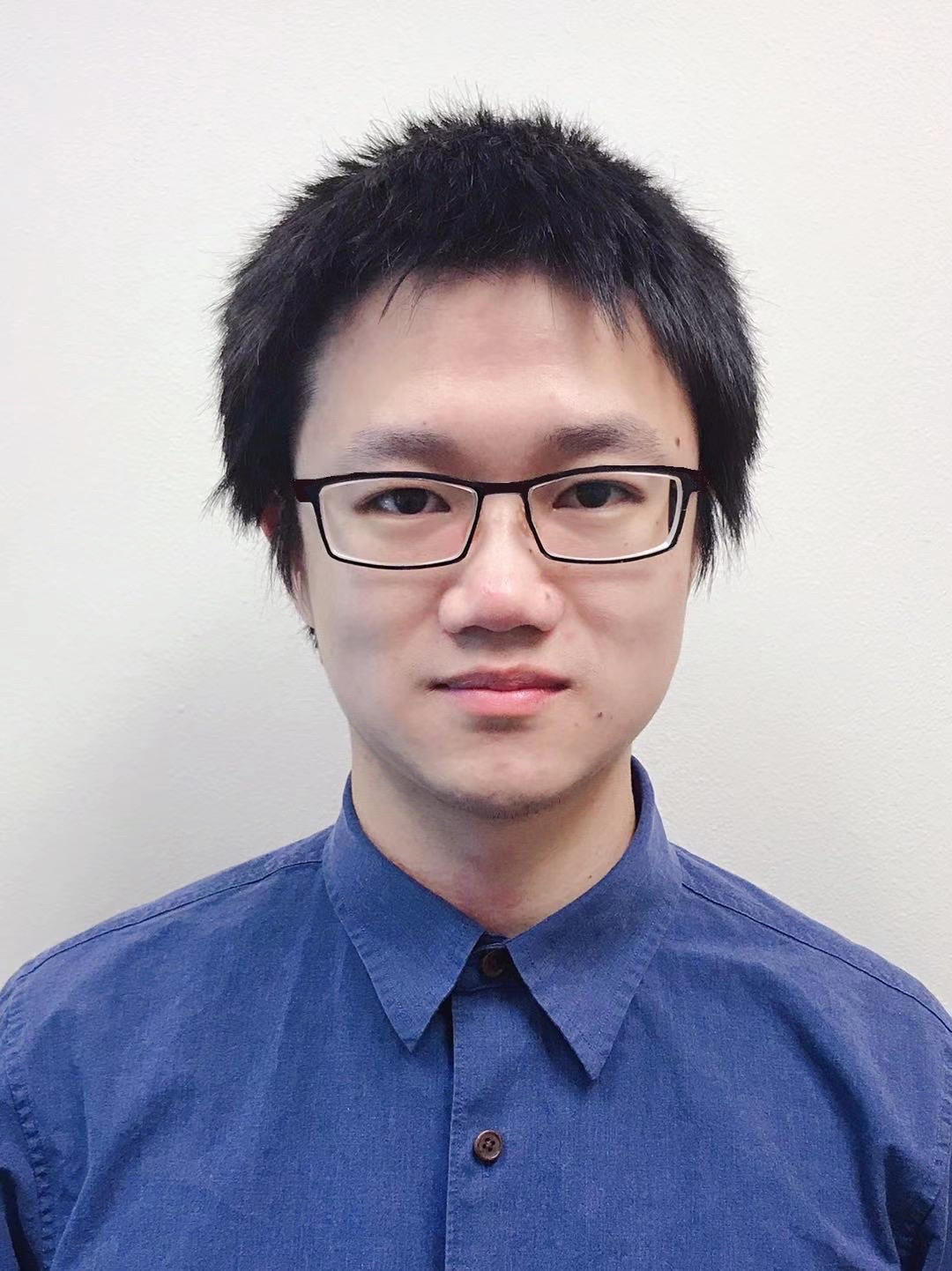}}]{Yebo Feng}
    is a Ph.D. candidate in the Department of Computer and Information Science at the University of Oregon (UO),
    where he conducts his research in the Center for Cyber Security and Privacy.
    He received his M.S. degree from UO in 2018, 
    B.E. from Yangzhou University in 2016, all in computer science.
    His research interests include computer security, anomaly detection, DDoS defense, and data analysis.
    He is the recipient of the Best Paper Award of 2019 IEEE CNS, Gurdeep Pall Graduate Student Fellowship of UO,
    and Ripple Research Fellowship.
    He has served as the reviewer of IEEE TDSC, IEEE TIFS, IEEE JSAC, and ACM TKDD.
    He was on the program committees of several international conferences, such as CYBER, SECURWARE, and B2C. He was also on the AE committees of USENIX OSDI and USENIX ATC.
\end{IEEEbiography}

\begin{IEEEbiography}[{\includegraphics[width=1in,height=1.25in,clip,keepaspectratio]{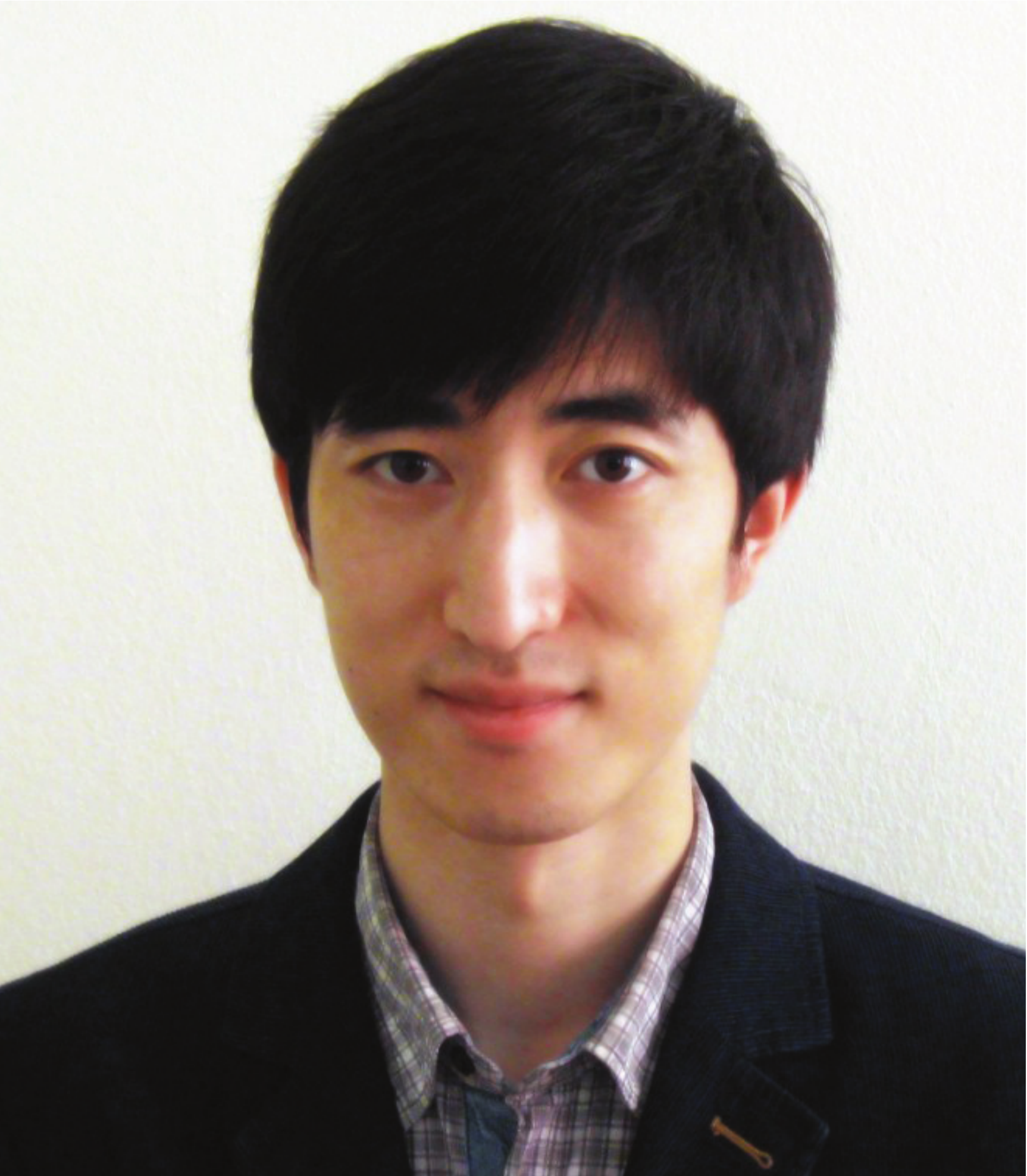}}]{Lei Jiao}
  received the Ph.D. degree in computer science from University of G\"ottingen, Germany.
  He is currently an assistant professor at the Department of Computer and Information Science, University of Oregon, USA.
  Previously he worked as a member of technical staff at Alcatel-Lucent/Nokia Bell Labs in Dublin, Ireland and also as a researcher at IBM Research in Beijing, China.
  He is interested in exploring optimization, control, learning, mechanism design, and game theory to manage and orchestrate large-scale distributed computing and communication infrastructures, services, and applications.
  He has published papers in journals such as JSAC, TON, TMC, and TPDS, and in conferences such as MOBIHOC, INFOCOM, ICNP, ICDCS, SECON, and IPDPS.
  He served as a guest editor for IEEE JSAC Series on Network Softwarization and Enablers.
  He was on the program committees of many conferences including MOBIHOC, INFOCOM, ICDCS, and IWQoS, and was the program chair of multiple workshops with INFOCOM and ICDCS.
  He was also a recipient of the Best Paper Awards of IEEE CNS 2019 and IEEE LANMAN 2013, and the 2016 Alcatel-Lucent Bell Labs UK and Ireland Recognition Award.
\end{IEEEbiography}

\begin{IEEEbiography}[{\includegraphics[width=1in,height=1.25in,clip,keepaspectratio]{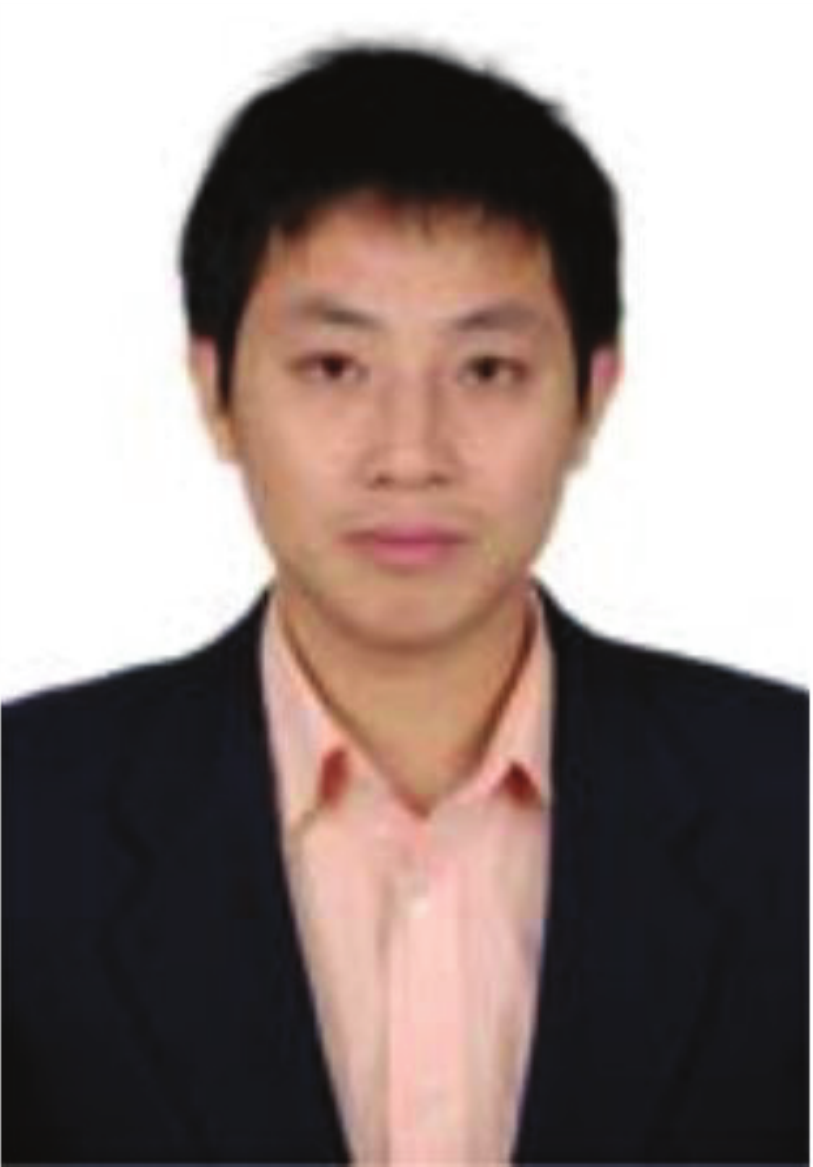}}]{Yulong~Shen}
  received the B.S. and M.S. degrees in computer science and PhD degree in cryptography from Xidian University, Xi’an, China, in 2002, 2005, and 2008, respectively. He is currently a Professor with the School of Computer Science and Technology, Xidian University, where he is also an Associate Director of the Shaanxi Key Laboratory of Network and System Security and a member of the State Key Laboratory of Integrated Services Networks. His research interests include wireless network security and cloud computing security. He has also served on the technical program committees of several international conferences, including ICEBE, INCoS, CIS, and SOWN.
\end{IEEEbiography}

\begin{IEEEbiography}[{\includegraphics[width=1in,height=1.25in,clip,keepaspectratio]{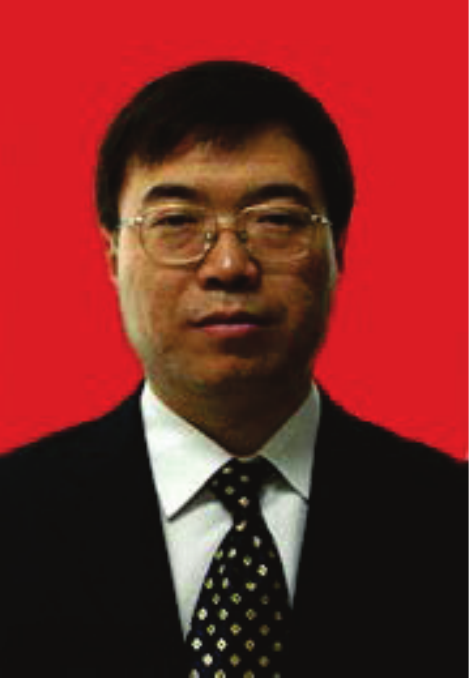}}]{Jianfeng~Ma}
 received the B.S. degree in computer science from Shaanxi Normal University in 1982, and M. S. degree in computer science from Xidian University in 1992, and the Ph. D. degree in computer science from Xidian University in 1995. Currently he is the directer of Department of Cyber engineering and a professor in School of Cyber Engineering, Xidian University. He has published over 150 journal and conference papers. His research interests include information security, cryptography, and network security.
\end{IEEEbiography}

\end{document}